\documentclass{article}
\usepackage[utf8]{inputenc}

\usepackage{amssymb}
\usepackage{amsmath}

\usepackage{braket}

\newtheorem{theorem}{Theorem}
\newtheorem{proposition}[theorem]{Proposition}

\usepackage[toc, page]{appendix}

\usepackage[nottoc]{tocbibind}
\usepackage[bookmarks=true]{hyperref}
\usepackage[numbered]{bookmark}

\hypersetup{
	colorlinks	= true,
	urlcolor	= blue,
	linkcolor	= blue,
	citecolor	= blue
}

\usepackage[margin=1.4in]{geometry}

\title{
	Self-adjointness of a simplified Dirac interaction operator without any cutoffs
}
\author{
	Mads J.\ Damgaard\footnote{
		B.Sc.\ at the Niels Bohr Institute, University of Copenhagen.
		E-mail: fxn318@alumni.ku.dk.
	}
}
\date{\today}

\begin{document}
\maketitle

\begin{abstract}

We show that a simplified version of the Dirac interaction operator given by $\hat H_\mathrm{I} \propto \int d\mathbf{k}d\mathbf{p}(\hat a(\mathbf{k}) + \hat a^\dagger(-\mathbf{k})) \hat b^\dagger(\mathbf{p} + \mathbf{k}) \hat b(\mathbf{p})/\sqrt{|\mathbf{k}|}$ is self-adjoint on a certain domain that is dense in the Hilbert space, even without any cutoffs. The technique that we use for showing this can potentially be extended to a much wider range of operators as well. This technique might therefore potentially lead to more mathematically well-defined theories of QFT in the future.

\end{abstract}

\section{Introduction}

In quantum field theory (QFT), one often encounters divergencies that need to be treated in some way. This sometimes means introducing certain cutoffs to the operators involved.
However, without the knowledge that there exist an underlying self-adjoint Hamiltonian for a given theory, it is not known whether the predictions of that theory will converge in the limit when these cutoffs are lifted.

In this paper, we will show that a certain interaction operator, which can be seen as a simplified version of the Dirac interaction operator, is self-adjoint on some domain of the Hilbert space. The hope is then that this technique can be extended to other operators as well, and in particular 
to ones that are more realistic as physical Hamiltonians.

\section{A simplified Dirac interaction operator}

The particular operator that we will investigate in this paper is a simplified version of the Dirac interaction operator, given formally by
\begin{equation}
\begin{aligned}
	\hat H_\mathrm{I} = \int \frac{d\mathbf{k}\,d\mathbf{p}}{(2\pi)^6}
		\frac{1}{\sqrt{|\mathbf{k}|}}
		\big(
			\hat a(\mathbf{k}) + \hat a^\dagger(-\mathbf{k})
		\big)
		\hat b^\dagger(\mathbf{p} + \mathbf{k}) \hat b(\mathbf{p}).
	\label{hat_H_v1}
\end{aligned}
\end{equation}
Here $\hat a^\dagger(\mathbf{k})$ and $\hat a(\mathbf{k})$ are the well-known creation and annihilation operators for photons and $\hat b^\dagger(\mathbf{k})$ and $\hat b(\mathbf{k})$ are the creation and annihilation operators for some fermions in the system. The operator $\hat H_\mathrm{I}$ thus either annihilates a photon with wave vector $\mathbf{k}$ or creates photon with wave vector $-\mathbf{k}$, with transition amplitude $1/\sqrt{\mathbf{k}}$, and then transitions a fermion in a state of momentum $\mathbf{p}$ to a state of momentum $\mathbf{p}+\mathbf{k}$.

The Hilbert space is thus a Fock space where the photon number can vary from state\footnote{
	We use the term `state' interchangeably with `vector in $\mathbf{H}$' in this paper, as is conventional in physics literature.
}
to state. The number of fermions is fixed, on the other hand, and for simplicity, we will assume that we have only a single fermion in the system.
This means that the Hilbert space, call it $\mathbf{H}$, will be isomorphic to the space, call it $\mathbf{H}'$, of all infinite sequences of square-integrable functions,
\begin{equation}
\begin{aligned}
	\psi = (\psi_0, \psi_1, \psi_2, \ldots),
\end{aligned}
\end{equation}
where $\psi_n : \mathbb{R}^{3n+3} \to \mathbb{C}$ for each $n\in\mathbb{N}_0$.
We can then take the first $3n$ coordinates in the domain of $\psi_n$ to represent the photon coordinates in momentum space, $\mathbf{k}_1, \ldots, \mathbf{k}_n$, and let the last 3 coordinates represent the momentum, $\mathbf{p}$, of the fermion. In a generalized momentum basis, we could thus write $\ket{\psi} \in \mathbf{H}$ as
\begin{equation}
\begin{aligned}
	\ket{\psi} &=
		 \ket{\psi_0} + \ket{\psi_1} + \ket{\psi_2} + \ldots \,=
		\int d\mathbf{p}\, \psi_0(\mathbf{p})\ket{\mathbf{p}} +
		\int d\mathbf{k}_1\,d\mathbf{p}\,
			\psi_1(\mathbf{k}_1; \mathbf{p})\ket{\mathbf{k}_1;\mathbf{p}}
	\\&\quad\quad\quad +
		\int d\mathbf{k}_1\,d\mathbf{k}_2\,d\mathbf{p}\,
			\psi_2(\mathbf{k}_1, \mathbf{k}_2; \mathbf{p})\ket{\mathbf{k}_1, 
			\mathbf{k}_2;\mathbf{p}}
		+
		\ldots,
\end{aligned}
\end{equation}
where we for any $\psi_n : \mathbb{R}^{3n+3} \to \mathbb{C}$ take $\ket{\psi_n}\in\mathbf{H}$ to denote the state given by
\begin{equation}
\begin{aligned}
	\ket{\psi_n} = \ket{(0, \ldots, 0, \psi_n, 0, \ldots)}.
\end{aligned}
\end{equation}
The inner product between any two states,
$\ket{\psi}$ and $\ket{\phi}$, is then naturally given by
\begin{equation}
\begin{aligned}
	\braket{\psi | \phi} =
		\sum_{n=0}^\infty \int 
		\psi_n(\mathbf{k}_1, \ldots, \mathbf{k}_n; \mathbf{p})^* 
		\phi_n(\mathbf{k}_1, \ldots, \mathbf{k}_n; \mathbf{p})
		\,d\mathbf{k}_1 \cdots d\mathbf{k}_n\,d\mathbf{p},
	\label{inner_product}
\end{aligned}
\end{equation}
and the norm of any state $\ket{\psi}$, which we will denote by $\|\psi\|$, is given by the relation
\begin{equation}
\begin{aligned}
	\|\psi\|^2 =
		\sum_{n=0}^\infty \int 
		|\psi_n(\mathbf{k}_1, \ldots, \mathbf{k}_n; \mathbf{p})|^2
		d\mathbf{k}_1 \cdots d\mathbf{k}_n\,d\mathbf{p}
	\equiv 
		\sum_{n=0}^\infty \|\psi_n\|^2.
	\label{norm}
\end{aligned}
\end{equation}

Since photons are bosons, we should also at some point like to require that all these $\psi_n$'s have to be symmetric with respect to permutations of the $\mathbf{k}$-parameters in order for $\ket{\psi}$ to belong to $\mathbf{H}$.
But, as is also briefly discussed in Appendix \ref{app_symmetrize}, it is in fact a simple matter of extending the proposition obtained in this paper to a symmetrized Hilbert space.

Now, to define $\hat H_\mathrm{I}$ more precisely such that it behaves in accordance with Eq.\ (\ref{hat_H_v1}), we can then define an operator, $\hat A$, given by 
\begin{equation}
\begin{aligned}
	\hat A \psi =
		\hat A^+ \psi + \hat A^- \psi
	=
		(
			\hat A^-_1 \psi_1,\;
			\hat A^+_0 \psi_0 + \hat A^-_2 \psi_2,\;
			\hat A^+_1 \psi_1 + \hat A^-_3 \psi_3,\;
			\hat A^+_2 \psi_2 + \hat A^-_4 \psi_4,\,
			\ldots
		)
	\label{hat_A}
\end{aligned}
\end{equation}
for all $\psi \in \mathbf{H}'$ (regardless of whether $\|\hat A \psi\|$ is finite or not), where we for all $n \in \mathbb{N}_+$ define
\begin{gather}
	\hat A^+_{n-1} \psi_{n-1}(\mathbf{k}_1, \ldots, \mathbf{k}_{n}; \mathbf{p}) =
		\frac{1}{\sqrt{n}}\sum_{i=1}^{n}
			\frac{1}{\sqrt{|\mathbf{k}_{i}|}}
			\psi_{n-1}(
				\mathbf{k}_1, \ldots, \widehat{\mathbf{k}_{i\,}}, \ldots, 
				\mathbf{k}_{n}; \mathbf{p} + \mathbf{k}_{i}
			),
	\label{A_plus_n_minus_1}\\
	\hat A^-_{n} \psi_{n}(\mathbf{k}_1, \ldots, \mathbf{k}_{n-1}; \mathbf{p}) =
		\frac{1}{\sqrt{n}}\sum_{i=1}^{n} \int
		\frac{1}{\sqrt{|\mathbf{k}_{n}|}}
		\psi_{n}(
			\mathbf{k}_1, \ldots, \mathbf{k}_{i-1}, \mathbf{k}_{n}, \mathbf{k}_{i}, \ldots,
			\mathbf{k}_{n-1}; \mathbf{p} - \mathbf{k}_n
		)
		\,d\mathbf{k}_{n}.
	\label{A_minus_n}
\end{gather}
Here we take $\mathbf{k}_1, \ldots, \widehat{\mathbf{k}_{i\,}}, \ldots, \mathbf{k}_{n}$ to denote the list of $\mathbf{k}_j$'s with $j$ going from 1 to $n$ but skipping $j=i$. In other words, $\mathbf{k}_1, \ldots, \widehat{\mathbf{k}_{i\,}}, \ldots, \mathbf{k}_{n}$ denotes the same as
$\mathbf{k}_1, \ldots, \mathbf{k}_{i-1}, \mathbf{k}_{i+1}, \ldots, \mathbf{k}_{n}$.
We will then define $\hat H_\mathrm{I}$ to be formally equivalent to $\hat A$, namely such that
\begin{equation}
\begin{aligned}
	\hat H_\mathrm{I} \ket{\psi} = \ket{ \hat A \psi}
	\label{hat_H_v2}
\end{aligned}
\end{equation}
for all $\ket{\psi} \in \mathrm{Dom}(\hat H_\mathrm{I})$. But unlike $\hat A$, we will not choose the domain of $\hat H_\mathrm{I}$ to be all of the Hilbert space. Rather we will let $\mathrm{Dom}(\hat H_\mathrm{I})$ be a proper subset of $\mathbf{H}$, which we will define in the following section.

\section[A domain on which the operator is self-adjoint]{A domain on which $\hat H_\mathrm{I}$ is self-adjoint} \label{sect_domain_and_prop}

We will now define a domain for $\hat H_\mathrm{I}$ on which we wish to show that this operator is self-adjoint.
%
First of all, define $p_{n} : \mathbb{R}^{3(n-1)} \to \mathbb{R}$ for all $n \in\mathbb{N}$, $n\geq 2$, by
\begin{equation}
\begin{aligned}
	p_{n}(\mathbf{k}_{1}, \ldots, \mathbf{k}_{n-1}) =
		e^{n} \prod_{i=1}^{n-1} (\mathbf{k}_{i}^2 + 1)^2
	\label{p_n},
\end{aligned}
\end{equation}
and define also $E_n$ by
\begin{equation}
\begin{aligned}
	E_n = \big\{
		(\mathbf{k}_{1}, \ldots, \mathbf{k}_{n})\in \mathbb{R}^{3n} \;\big|\;
			e^{ p_n(\mathbf{k}_1, \ldots, \mathbf{k}_{n-1})/2 }
			<
			|\mathbf{k}_{n}|
			<
			e^{ p_n(\mathbf{k}_1, \ldots, \mathbf{k}_{n-1}) }
	\big\}.
	\label{E_n}
\end{aligned}
\end{equation}
Note that with this definition, we have $|\mathbf{k}_n| > \max(|\mathbf{k}_{1}|, \ldots, |\mathbf{k}_{n-1}|)$ for all $n\geq 2$
whenever $(\mathbf{k}_{1}, \ldots, \mathbf{k}_{n}) \in E_n$.
Then define $F_n$ recursively for all $n\geq 2$ by
\begin{equation}
\begin{aligned}
	F_{n} = \mathcal{S} \big( (F_{n-1}^\complement \times \mathbb{R}^3) \cap E_{n} \big),
	\quad
	F_{1}^\complement = \mathbb{R}^3,
	\label{F_n}
\end{aligned}
\end{equation}
where we take $\mathcal{S}$ to denote a symmetrizing operator for sets, such that for any set $X$ that is a subset of $\mathbb{R}^{3N}$ for some $N$, we take $\mathcal{S}X$ to be the set of all $(\mathbf x_1, \ldots, \mathbf x_{N}) \in \mathbb{R}^{3N}$ for which a permutation of $(\mathbf x_1, \ldots, \mathbf x_{N})$ exists that is a member of $X$.
Note also that we take $X^\complement$ to denote the complement of $X$.

Let then $T$ and $T^\complement$ be the two sequences of sets given by
\begin{equation}
\begin{aligned}
	T &=
		(
			\emptyset,\,
			F_1 \times \mathbb{R}^3,\,
			F_2 \times \mathbb{R}^3,\,
			F_3 \times \mathbb{R}^3,\,
			\ldots\,
		),
	\\
	T^\complement &=
		(
			\mathbb{R}^3,\,
			F_1^\complement \times \mathbb{R}^3,\,
			F_2^\complement \times \mathbb{R}^3,\,
			F_3^\complement \times \mathbb{R}^3,\,
			\ldots\,
		).
\end{aligned}
\end{equation}
And let furthermore $\hat 1_{T}$ and $\hat 1_{T^\complement}$ denote the two operators given for all $\psi \in \mathbf{H}'$ by
\begin{equation}
\begin{aligned}
	\hat 1_{T} \psi &=
		(
			1_{\emptyset} \psi_{0},\,
			1_{F_{1} \times \mathbb{R}^3} \psi_{1},\,
			1_{F_{2} \times \mathbb{R}^3} \psi_{2},\,
			1_{F_{3} \times \mathbb{R}^3} \psi_{3},\,
			\ldots\,
		),
	\\
	\hat 1_{T^\complement} \psi &=
		(
			1_{\mathbb{R}^3} \psi_{0},\,
			1_{F_{1}^\complement \times \mathbb{R}^3} \psi_{1},\,
			1_{F_{2}^\complement \times \mathbb{R}^3} \psi_{2},\,
			1_{F_{3}^\complement \times \mathbb{R}^3} \psi_{3},\,
			\ldots\,
		),
	\label{hat_1_Ts}
\end{aligned}
\end{equation}
where we for any set, $X$, we take $1_{X}$ to denote the indicator function of $X$.
Note also that we take the product of any two functions, $f$ and $g$ (which could be e.g.\ $1_{F_{n} \times \mathbb{R}^3}$ and $\psi_{n}$ in this case), to denote the function given by $x \mapsto f(x)g(x)$, 
$x \in \mathrm{Dom}(f) \cap \mathrm{Dom}(g)$.

With these definitions, we are now ready to define two subsets of $\mathbf{H}$, the intersection of which we will choose as our domain for $\hat H_\mathrm{I}$.
Let first $V$ be the set of all $\ket{\psi} \in \mathbf{H}$ where
\begin{equation}
\begin{aligned}
	\| \hat 1_{T} \hat A^+ \hat 1_{T^\complement} \psi \| < \infty.
	\label{V_def}
\end{aligned}
\end{equation}
And
let $U$ be the set of all $\ket{\psi} \in \mathbf{H}$ where
\begin{equation}
\begin{aligned}
	\ket{\hat A \psi} \in \mathbf{H},
\end{aligned}
\end{equation}
which is equivalent of saying that $\|\hat H_\mathrm{I} \ket{\psi}\!\| = \|\hat A \psi\| < \infty$.
We will then define $\mathrm{Dom}(\hat H_\mathrm{I})$ as
\begin{equation}
\begin{aligned}
	\mathrm{Dom}(\hat H_\mathrm{I}) = V \cap U.
\end{aligned}
\end{equation}

The proposition that we want to show in this paper then reads as follows.

\begin{proposition}
	Suppose that $\hat H_\mathrm{I}$ and $\mathrm{Dom}(\hat H_\mathrm{I})$ are defined as above. Then $\mathrm{Dom}(\hat H_\mathrm{I})$ is dense in $\mathbf{H}$ and $\hat H_\mathrm{I}$ is self-adjoint.
	\label{prop_1}
\end{proposition}

In order to show this in the following sections, we will start by introducing a set of vectors, $W$, which we hope to show is a subset of both $V$ and $U$. And we also hope to show that this $W$ spans a dense subspace of $\mathbf{H}$, giving us the first part of Proposition \ref{prop_1}. Then we want to show that $\hat H_\mathrm{I}$ is symmetric on its domain. And finally we want to show that the domain of its adjoint, $\mathrm{Dom}(\hat H_\mathrm{I}^*)$, is equal to $\mathrm{Dom}(\hat H_\mathrm{I})$, giving us the last part of Proposition \ref{prop_1}.

\section[Introducing W ⊂ V]{Introducing a set $W \subset V$} \label{sect_W_subset_V}

We will now construct a certain set, $W \subset \mathbf{H}$, which we will then show is a subset of $V$.
Let this $W$ be the set of all $\ket{\chi} \in \mathbf{H}$ of the form
\begin{equation}
\begin{aligned}
	\ket{\chi} = \ket{\chi_{m}} + \ket{\chi_{m+2}} + \ket{\chi_{m+4}} + \ldots,
	\quad
	m \in \mathbb{N}_0,
\end{aligned}
\end{equation}
for which we first of all require that for all $n\in\{m+2, m+4, m+6, \ldots\}$, we have
\begin{equation}
\begin{aligned}
	\chi_{n}(\mathbf{k}_1, \ldots, \mathbf{k}_{n}; \mathbf{p}) &=
		\frac{
			-\sqrt{n}\, 1_{E_{n}}(\mathbf{k}_{1}, \ldots, \mathbf{k}_{n})
			1_{D_{n-1}^\complement}(\mathbf{k}_1, \ldots, \mathbf{k}_{n-1})
		}{
			2\pi\sqrt{n-1}\,
			p_n(\mathbf{k}_1, \ldots, \mathbf{k}_{n-1})
			\sqrt{|\mathbf{k}_{n}|^5}
		}
	\\&\phantom{=}\quad\quad\quad
		\sum_{j=1}^{n-1} \frac{
			1
		}{
			\sqrt{|\mathbf{k}_{j}|}
		}\,
		\chi_{n-2}(
			\mathbf{k}_1, \ldots, \widehat{\mathbf{k}_{j}}, \ldots, \mathbf{k}_{n-1};
			\mathbf{p} + \mathbf{k}_{j} + \mathbf{k}_n
		),
	\label{chi_n}
\end{aligned}
\end{equation}
where $D_{m+1}$ is some superset of $F_{m+1}$ for which $D_{m+1} = \mathcal{S} D_{m+1}$,
and where for all $n > m + 2$, $D_{n-1}^\complement = \mathbb{R}^{3(n-1)}$, which means that $1_{D_{n-1}^\complement}$ is just 1 everywhere.
Furthermore, we require that
the functions $\hat A^-_m \chi_{m}$ and $1_{D_{m+1} \times \mathbb{R}^3}\hat A^+_m \chi_{m}$, and of course also $\chi_{m}$, are all
square-integrable.

We want to show that $\ket{\chi} \in V$ for all $\ket{\chi} \in W$.
In order to do this, we first of all note that for any $(\mathbf{k}_1, \ldots, \mathbf{k}_n; \mathbf{p})$ in the support of $\chi_n$, $n > m$, there exist a permutation of $(\mathbf{k}_1, \ldots, \mathbf{k}_{n-1})$, call it $(\mathbf{k}_1', \ldots, \mathbf{k}_{n-1}')$, such that
\begin{gather}
	(\mathbf{k}_1', \ldots, \mathbf{k}_{n-2}'; \mathbf{p} + \mathbf{k}_{n-1}' + \mathbf{k}_{n})
	\in
		\mathrm{supp}(\chi_{n-2}),
	\label{support_on_supp_chi_n_minus_2}
	\\
	(\mathbf{k}_1', \ldots, \mathbf{k}_{n-1}') \in
		D_{n-1}^\complement,
	\label{support_on_D}
	\\
	(\mathbf{k}_1', \ldots, \mathbf{k}_{n-1}', \mathbf{k}_{n}) \in
		E_{n},
	\label{support_on_E}
\end{gather}
where $\mathrm{supp}(\chi_{n-2})$ denotes the support of $\chi_{n-2}$.
Here we have thus also used the fact that both $1_{D_{n-1}^\complement}(\mathbf{k}_1, \ldots, \mathbf{k}_{n-1})$ and $1_{E_{n}}(\mathbf{k}_{1}, \ldots, \mathbf{k}_{n})$ are invariant with respect to permutations of $\mathbf{k}_1, \ldots, \mathbf{k}_{n-1}$.

Now, for $n = m+2$, Eq.\ (\ref{support_on_D}) gives us that $(\mathbf{k}_1', \ldots, \mathbf{k}_{n-1}') \in D_{m+1}^\complement \subset F_{n-1}^\complement$. Using Eq.\ (\ref{support_on_E}) as well, we thus get that $(\mathbf{k}_1', \ldots, \mathbf{k}_{n-1}', \mathbf{k}_{n}) \in (F_{n-1}^\complement \times \mathbb{R}^3) \cap E_n \subset F_n$ in this case. And since $F_n = \mathcal{S} F_n$, we therefore also get that $(\mathbf{k}_1, \ldots, \mathbf{k}_{n}) \in F_n$. So the support of $\chi_{m+2}$ must be a subset of $F_n \times \mathbb{R}^3$, $n = m+2$.

Then for the cases when $n > m+2$, if we assume that $\mathrm{supp}(\chi_{n-2}) \subset F_{n-2} \times \mathbb{R}^3$, we get from Eqs. (\ref{support_on_supp_chi_n_minus_2}) and (\ref{support_on_E}) that $(\mathbf{k}_1', \ldots, \mathbf{k}_{n-1}', \mathbf{k}_{n}) \in (F_{n-2} \times \mathbb{R}^6) \cap E_n$. If we can then show that
\begin{equation}
	F_{n-1} \times \mathbb{R}^3 \subset F_{n}^\complement
	\label{F_n_minus_1_productions_are_in_F_n_complement}
\end{equation}
for all $n \geq 3$, we will thus get that $(F_{n-2} \times \mathbb{R}^6) \cap E_n \subset (F_{n-1}^\complement \times \mathbb{R}^3) \cap E_n \subset F_n$ for all $n \geq 4$, which will tell us that $(\mathbf{k}_1, \ldots, \mathbf{k}_{n}) \in \mathcal{S} F_n = F_n$.
And by recursion, this will then yield us that $\chi_n$ is supported only on $F_n \times \mathbb{R}^3$ for all $n > m$.

To show this lemma of Eq.\ (\ref{F_n_minus_1_productions_are_in_F_n_complement}), let us assume that there exist some $(\mathbf{k}_1, \ldots, \mathbf{k}_n)$ that is a member of both $(F_{n-1} \times \mathbb{R}^3)$ and $F_n$ at the same time and look for a contradiction.
This implies that there must exist some permutations, $(\mathbf{k}_1', \ldots, \mathbf{k}_{n}')$ and $(\mathbf{k}_1'', \ldots, \mathbf{k}_{n}'')$, of $(\mathbf{k}_1, \ldots, \mathbf{k}_{n})$ such that
\begin{equation}
	(\mathbf{k}_1', \ldots, \mathbf{k}_{n-1}') \in 
		(F_{n-2}^\complement \times \mathbb{R}^3) \cap E_{n-1},
	\quad
	(\mathbf{k}_1'', \ldots, \mathbf{k}_{n}'') \in
		(F_{n-1}^\complement \times \mathbb{R}^3) \cap E_{n}.
	\label{lemma_permutations_eq}
\end{equation}
Recalling Eqs.\ (\ref{p_n}--\ref{E_n}), we see that $p_{n}(\mathbf{k}_1, \ldots, \mathbf{k}_{n-1})/2 > p_{n-1}(\mathbf{k}_1, \ldots, \mathbf{k}_{n-2})$ for all $n\geq 3$. And from this we see that $|\mathbf{k}_{n}''|$ is greater than $|\mathbf{k}_{n-1}'|$, which is again greater than all other $|\mathbf{k}_1'|, \ldots, |\mathbf{k}_{n-2}'|$. Therefore, $\mathbf{k}_{n}''$ cannot be contained in $(\mathbf{k}_1', \ldots, \mathbf{k}_{n-1}')$. The tuples $(\mathbf{k}_1', \ldots, \mathbf{k}_{n-1}')$ and $(\mathbf{k}_1'', \ldots, \mathbf{k}_{n-1}'')$ must therefore be permutations of each other. And since $(\mathbf{k}_1', \ldots, \mathbf{k}_{n-1}') \in F_{n-1}$, we thus get that $(\mathbf{k}_1'', \ldots, \mathbf{k}_{n-1}'') \in \mathcal{S} F_{n-1} = F_{n-1}$. But on the other hand, we also have from Eq.\ (\ref{lemma_permutations_eq}) that $(\mathbf{k}_1'', \ldots, \mathbf{k}_{n-1}'') \in F_{n-1}^\complement$. We thus get the contradiction that we sought after, which shows that no such tuple as $(\mathbf{k}_1, \ldots, \mathbf{k}_n)$ exists. This means that $F_{n-1} \times \mathbb{R}^3$ and $F_n$ must be disjoint, which means that Eq.\ (\ref{F_n_minus_1_productions_are_in_F_n_complement}) is true for all $n \geq 3$.

We have thus obtained that $\chi_n$ is supported only on $F_n \times \mathbb{R}^3$ for all $n > m$. We then want to use this result to show that $\| \hat 1_{T} \hat A^+ \hat 1_{T^\complement} \chi \| < \infty$ for all $\ket{\chi} \in W$, which, according to the definition of $V$ in Eq.\ (\ref{V_def}) will imply that $W \subset V$. We then first of all see that our result gives us that
\begin{equation}
	\ket{\hat 1_{T^\complement} \chi} = \ket{1_{F_m^\complement \times \mathbb{R}^3}\chi_m}.
\end{equation}
And since it follows that
\begin{equation}
	\| \hat 1_{T} \hat A^+ \hat 1_{T^\complement} \chi \| =
		\|
			1_{F_{m+1} \times \mathbb{R}^3} \hat A^+_m (
				1_{F_m^\complement \times \mathbb{R}^3} \chi_m
			)
		\|
	\leq
		\| 1_{D_{m+1} \times \mathbb{R}^3}\hat A^+_m \chi_m \|,
\end{equation}
we thus, by our assumption that $1_{D_{m+1} \times \mathbb{R}^3}\hat A^+_m \chi_m$ is square-integrable, get that $\| \hat 1_{T} \hat A^+ \hat 1_{T^\complement} \chi \| < \infty$ for all $\ket{\chi} \in W$.
We can therefore conclude that $W \subset V$.

Before we move on to show that $W \subset U$ in the following section, note first that $W$ might actually be empty for all we know at this point, since there might not exist any $\ket{\chi}$ of the above form for which $\|\chi\| < \infty$, which is required in order for $\ket{\chi}$ to be a member of $\mathbf{H}$. But as we will show in the following two sections, $\|\chi\|$ is in fact always finite whenever $\|\chi_m\|$ is. 

\section[Showing that W ⊂ U]{Showing that $W$ is a subset of $U$} \label{sect_W_subset_U}


In order to show that $W \subset U$, we need to show that $\|\hat A \chi\| < \infty$ for all $\ket{\chi} \in W$.
Let us start by dividing $\hat A^-_n$, for each $n\in\mathbb{N}_+$, up into $n$ terms, namely such that $\hat A^-_n = \sum_{l=1}^{n} \hat A^-_{n,l}$ with
\begin{equation}
\begin{aligned}
	\hat A^-_{n,l} \psi_{n}(
		\mathbf{k}_1, \ldots, \widehat{\mathbf{k}_{l\,}}, \ldots, \mathbf{k}_{n};
		\mathbf{p}
	) =
		\frac{1}{\sqrt{n}} \int
			\frac{1}{\sqrt{|\mathbf{k}_{l}|}}
			\psi_{n}(
				\mathbf{k}_1, \ldots, \mathbf{k}_{n}; \mathbf{p} - \mathbf{k}_l
			)
			\,d\mathbf{k}_{l}
	\label{hat_A_n_l}
\end{aligned}
\end{equation}
for all $\psi \in \mathbf{H}'$. 
Now, one of the main ideas behind how we constructed the formula for $\chi_n$ above is that the contribution to $(\hat A \chi_n)_{n-1} = \hat A^+_{n-2} \chi_{n-2} + \hat A^-_{n} \chi_{n}$ coming from $\hat A^{-}_{n,n} \chi_{n}$ 
is designed to
cancel enough of the otherwise infinite-normed $\hat A^+_{n-2} \chi_{n-2}$ to give us
$\|\hat A^-_{n,n} \chi_{n} + \hat A^+_{n-2} \chi_{n-2}\| < \infty$.
Let us start by confirming that this is indeed the case.

From Eqs.\ (\ref{hat_A_n_l}) and (\ref{chi_n}), we get that
\begin{equation}
\begin{aligned}
	\hat A^-_{n,n} \chi_{n}(
		\mathbf{k}_1, \ldots, \mathbf{k}_{n-1}; \mathbf{p}
	) &=
		\frac{1}{\sqrt{n}} \int
		\frac{1}{\sqrt{|\mathbf{k}_{n}|}}
		\chi_{n}(
			\mathbf{k}_1, \ldots, \mathbf{k}_{n}; \mathbf{p} - \mathbf{k}_n
		)
		\,d\mathbf{k}_{n}
	\\&=
		\frac{
			-1_{D_{n-1}^\complement}(\mathbf{k}_1, \ldots, \mathbf{k}_{n-1})
		}{
			2\pi\sqrt{n-1}\,
			p_n(\mathbf{k}_1, \ldots, \mathbf{k}_{n-1})
		}
		\int
			\frac{
				1_{E_{n}}(\mathbf{k}_{1}, \ldots, \mathbf{k}_{n})
			}{
				|\mathbf{k}_{n}|^3
			}
			\,d\mathbf{k}_n
	\\&\phantom{=}\quad\quad\quad
		\sum_{j=1}^{n-1} \frac{
			1
		}{
			\sqrt{|\mathbf{k}_{j}|}
		}\,
		\chi_{n-2}(
			\mathbf{k}_1, \ldots, \widehat{\mathbf{k}_{j}}, \ldots, \mathbf{k}_{n-1};
			\mathbf{p} + \mathbf{k}_{j}
		).
\end{aligned}
\end{equation}
Recalling the definition of $E_n$ in Eq.\ (\ref{E_n}), we can then evaluate the integral over $\mathbf{k}_n$ using polar coordinates, giving us
\begin{equation}
\begin{aligned}
	\int
		\frac{1_{E_{n}}(\mathbf{k}_{1}, \ldots, \mathbf{k}_{n})}{|\mathbf{k}_n|^3}
		\,d\mathbf{k}_{n}
	=
		\int_{
				e^{p_n(\mathbf{k}_1, \ldots, \mathbf{k}_{n-1})/2}
			}^{
				e^{p_n(\mathbf{k}_1, \ldots, \mathbf{k}_{n-1})}
			}
			4\pi k^{-1}
			\,dk
	=
		2\pi p_n(\mathbf{k}_1, \ldots, \mathbf{k}_{n-1}).
	\label{E_n_over_k_n_to_the_3rd_integral}
\end{aligned}
\end{equation}
We thus get that
\begin{equation}
\begin{aligned}
	\hat A^-_{n,n} \chi_{n}(\mathbf{k}_1, \ldots, \mathbf{k}_{n-1}; \mathbf{p}) &= 
		\sum_{j=1}^{n-1}
		\frac{-1_{D_{n-1}^\complement}(\mathbf{k}_1, \ldots, \mathbf{k}_{n-1}) }{
			\sqrt{n-1} \sqrt{|\mathbf{k}_{j}|}
		}
		\chi_{n-2}(
			\mathbf{k}_1, \ldots, \widehat{\mathbf{k}_{j}}, \ldots, \mathbf{k}_{n-1};
			\mathbf{p} + \mathbf{k}_{j}
		).
\end{aligned}
\end{equation}
Let us then compare this to $\hat A^+_{n-2} \chi_{n-2}$, which, as can be seen from Eq.\ (\ref{A_plus_n_minus_1}), is given by
\begin{equation}
\begin{aligned}
	\hat A^+_{n-2} \chi_{n-2}(\mathbf{k}_1, \ldots, \mathbf{k}_{n-1}; \mathbf{p}) =
		\sum_{j=1}^{n-1}
		\frac{1}{\sqrt{n-1} \sqrt{|\mathbf{k}_{j}|}}
		\chi_{n-2}(
			\mathbf{k}_1, \ldots, \widehat{\mathbf{k}_{j}}, \ldots, 
			\mathbf{k}_{n-1}; \mathbf{p} + \mathbf{k}_{j}
		).
	\label{A_plus_n_minus_2}
\end{aligned}
\end{equation}
Using the fact that $1-1_{D_{n-1}^\complement} = 1_{D_{n-1}}$, we therefore obtain that
\begin{equation}
\begin{aligned}
	&(\hat A^-_{n,n} \chi_{n} + \hat A^+_{n-2} \chi_{n-2})(
		\mathbf{k}_1, \ldots, \mathbf{k}_{n-1}; \mathbf{p}
	)
	\\&\quad\quad\quad=
		\frac{
			1_{D_{n-1}}(
				\mathbf{k}_1, \ldots, \mathbf{k}_{n-1}
			)
		}{
			\sqrt{n-1}
		}
		\sum_{j=1}^{n-1} \frac{
			1
		}{
			\sqrt{|\mathbf{k}_{j}|}
		}
		\chi_{n-2}(
			\mathbf{k}_1, \ldots, \widehat{\mathbf{k}_{j}}, \ldots, \mathbf{k}_{n-1};
			\mathbf{p} + \mathbf{k}_{j}
		)
	\\&\quad\quad\quad =
		(1_{D_{n-1} \times \mathbb{R}^3}\hat A^+_{n-2} \chi_{n-2})(
			\mathbf{k}_1, \ldots, \mathbf{k}_{n-1}; \mathbf{p}
		),
\end{aligned}
\end{equation}
where have used Eq.\ (\ref{A_plus_n_minus_2}) again for the last equality to recognize the expression as simply the formula for
$1_{D_{n-1} \times \mathbb{R}^3}\hat A^+_{n-2} \chi_{n-2}$.
This shows us that
\begin{equation}
\begin{aligned}
	\hat A^-_{n,n} \chi_{n} + \hat A^+_{n-2} \chi_{n-2} =
		1_{D_{n-1} \times \mathbb{R}^3}\hat A^+_{n-2} \chi_{n-2}
\end{aligned}
\end{equation}
for all $n \geq m+2$.
And since $D_{n-1}$ is just the empty set for all $n \geq m+4$, we thus get that
\begin{equation}
\begin{aligned}
	\sum_{n=m+2}^{\infty}\|\hat A^-_{n,n} \chi_{n} + \hat A^+_{n-2} \chi_{n-2}\|^2 =
		\|1_{D_{m+1} \times \mathbb{R}^3}\hat A^+_m \chi_{m}\|^2,
	\label{that_term}
\end{aligned}
\end{equation}
which is finite by assumption.

This result seems promising. It shows that, despite the fact that $\|\hat A^+_{n} \psi_n \|$ is always infinite for any $\psi \in \mathbf{H}'$,
which is not hard to see,
whereas $\|\hat A^-_{n} \psi_n \|$ is often finite, there are in fact some $\psi \in \mathbf{H}'$ for which $\|\hat A^-_{n} \psi_n \|$ is infinite as well. And it seems that we might be able to use these kinds of solutions to cancel the infinite parts of $\hat A \psi$ that comes from $\sum_n \hat A^+_{n} \psi_n$ in order to get a finite $\|\hat A \psi \|$.
However, we have not quite shown this yet since we still need to confirm that the other contributions to $\hat A^-_{n} \chi_n$, namely the $\hat A^-_{n,l} \chi_n$-contributions with $l \neq n$, does not also have infinite norms.

In order to do this, let us first split $\chi_{n}$ up into $n-1$ terms such that $\chi_{n} = \sum_{j=1}^{n-1} \chi_{n,j}$, with $\chi_{n,j}$ defined by
\begin{equation}
\begin{aligned}
	\chi_{n,j}(\mathbf{k}_1, \ldots, \mathbf{k}_{n}; \mathbf{p}) &=
		\frac{
			-e^{-n} \sqrt{n}\, 1_{E_{n}}(\mathbf{k}_{1}, \ldots, \mathbf{k}_{n})
			1_{D_{n-1}^\complement}(\mathbf{k}_1, \ldots, \mathbf{k}_{n-1})
		}{
			2\pi\sqrt{n-1}
			\prod_{i=1}^{n-1} (\mathbf{k}_{i}^2 + 1)^2
			\sqrt{|\mathbf{k}_{n}|^5}
		}
	\\&\phantom{=}\quad\quad\quad
		\frac{
			1
		}{
			\sqrt{|\mathbf{k}_{j}|}
		}\,
		\chi_{n-2}(
			\mathbf{k}_1, \ldots, \widehat{\mathbf{k}_{j}}, \ldots, \mathbf{k}_{n-1};
			\mathbf{p} + \mathbf{k}_{j} + \mathbf{k}_n
		).
	\label{chi_n_j_v1}
\end{aligned}
\end{equation}
Here we have thus also substituted $p_{n}(\mathbf{k}_{1}, \ldots, \mathbf{k}_{n-1}) = \exp(n) \prod_{i=1}^{n-1} (\mathbf{k}_{i}^2 + 1)^2$.
We can then also note that $E_n \subset \mathbb{R}^{3n-3} \times B_{1,3}^\complement$ for all 
$n \geq 2$,
where we take $B_{R,N}$ to denote the origin-centered ball in $\mathbb{R}^N$ with a radius of $R$. This implies that $1_{E_n}(\mathbf{k}_{1}, \ldots, \mathbf{k}_{n}) = 1_{E_n}(\mathbf{k}_{1}, \ldots, \mathbf{k}_{n}) 1_{B_{1,3}^\complement}(\mathbf{k}_{n})$.
And if we also define $f_n$ for all $n\geq 2$ by
\begin{equation}
\begin{aligned}
	f_{n}(\mathbf{k}_1, \ldots, \mathbf{k}_{n}) =
		\frac{
			\sqrt{n}\, 1_{E_{n}}(\mathbf{k}_{1}, \ldots, \mathbf{k}_{n})
			1_{D_{n-1}^\complement}(\mathbf{k}_1, \ldots, \mathbf{k}_{n-1})
		}{
			2\pi\sqrt{n-1}
		},
	\label{f_n}
\end{aligned}
\end{equation}
we can therefore rewrite $\chi_{n,j}$ as
\begin{equation}
\begin{aligned}
	\chi_{n,j}(\mathbf{k}_1, \ldots, \mathbf{k}_{n}; \mathbf{p}) &=
		-e^{-n}
		\frac{
			1_{B_{1,3}^\complement}(\mathbf{k}_{n})
		}{
			\sqrt{|\mathbf{k}_{n}|^5}
		}
		\frac{
			f_{n}(\mathbf{k}_1, \ldots, \mathbf{k}_{n})
		}{
			(\mathbf{k}_{j}^2 + 1)^2 \sqrt{|\mathbf{k}_{j}|}\,
		}
		\frac{
			1
		}{
			\prod_{i=1, i\neq j}^{n-1} (\mathbf{k}_{i}^2 + 1)^2
		}
	\\&\phantom{=}\quad\quad\quad
		\chi_{n-2}(
			\mathbf{k}_1, \ldots, \widehat{\mathbf{k}_{j}}, \ldots, \mathbf{k}_{n-1};
			\mathbf{p} + \mathbf{k}_{j} + \mathbf{k}_n
		).
	\label{chi_n_j_v2}
\end{aligned}
\end{equation}
For $\|\hat A^-_{n,l} \chi_{n,j}\|^2$ with $l < n$, we thus obtain that
\begin{equation}
\begin{aligned}
	\|\hat A^-_{n,l} \chi_{n,j}\|^2 &=
		\frac{e^{-2n}}{n}
		\int d\mathbf{k}_n\,
			\frac{
				1_{B_{1,3}^\complement}(\mathbf{k}_{n})
			}{
				|\mathbf{k}_{n}|^5
			}
		\int d\mathbf{k}_1 \cdots \widehat{d\mathbf{k}_{l}} \cdots d\mathbf{k}_{n-1}\,
			d\mathbf{p}\,
		\bigg|
			\int d\mathbf{k}_l
			\frac{
				1
			}{
				\sqrt{\mathbf{k}_l}\,
			}
			\frac{
				1
			}{
				(\mathbf{k}_{j}^2 + 1)^2 \sqrt{|\mathbf{k}_{j}|}\,
			}
	\\&\phantom{=}\quad\quad\quad
			\frac{
				f_{n}(\mathbf{k}_1, \ldots, \mathbf{k}_{n})
			}{
				\prod_{i=1, i\neq j}^{n-1} (\mathbf{k}_{i}^2 + 1)^2
			}
			\chi_{n-2}(
				\mathbf{k}_1, \ldots, \widehat{\mathbf{k}_{j}}, \ldots, \mathbf{k}_{n-1};
				\mathbf{p} + \mathbf{k}_{j} - \mathbf{k}_l + \mathbf{k}_n
			)
		\bigg|^2.
	\label{A_l_chi_n_j_norm_squared_v1}
\end{aligned}
\end{equation}

In order to then derive a bound from Eq.\ (\ref{A_l_chi_n_j_norm_squared_v1}), we will first need a bound on $\|\chi_n\|^2$ for all $n$. From Eq.\ (\ref{chi_n_j_v2}), we get that
\begin{equation}
\begin{aligned}
	\|\chi_{n,j}\|^2 &=
		\frac{e^{-2n}}{n}
		\int d\mathbf{k}_n\,
			\frac{
				1_{B_{1,3}^\complement}(\mathbf{k}_{n})
			}{
				|\mathbf{k}_{n}|^5
			}
		\int d\mathbf{k}_j\,
			\frac{
				1
			}{
				(\mathbf{k}_{j}^2 + 1)^4 |\mathbf{k}_{j}|
			}
	\\&\phantom{=}\quad\quad\quad
		\int d\mathbf{k}_1 \cdots \widehat{d\mathbf{k}_{j}} \cdots d\mathbf{k}_{n-1}\,
			d\mathbf{p}\,
		\frac{
			\big|\chi_{n-2}(
				\mathbf{k}_1, \ldots, \widehat{\mathbf{k}_{j}}, \ldots, \mathbf{k}_{n-1};
				\mathbf{p} + \mathbf{k}_{j} + \mathbf{k}_n
			)\big|^2
		}{
			\big|f_{n}(\mathbf{k}_1, \ldots, \mathbf{k}_{n})\big|^{-2}
			\prod_{i=1, i\neq j}^{n-1} (\mathbf{k}_{i}^2 + 1)^4
		}.
	\label{chi_n_j_norm_v1}
\end{aligned}
\end{equation}
We can then first of all shift $\mathbf{p}$ by $-\mathbf{k}_j - \mathbf{k}_n$ for the integral. We can also note that both $|f_{n}(\mathbf{k}_1, \ldots, \mathbf{k}_{n})|^{-2}$ and $\prod_{i=1, i\neq j}^{n-1} (\mathbf{k}_{i}^2 + 1)^4$ are greater than or equal to 1 everywhere. And since the numerator of the integrand is real and non-negative everywhere, the integral can only increase if we remove this denominator. When we then integrate over $\mathbf{k}_1, \ldots, \widehat{\mathbf{k}_{j}}, \ldots, \mathbf{k}_{n-1}, \mathbf{p}$, we see that we simply get a factor of $\|\chi_{n-2}\|^2$. In other words, we get that
\begin{equation}
\begin{aligned}
	\|\chi_{n,j}\|^2 \leq
		\frac{e^{-2n}}{n}
		\|\chi_{n-2}\|^2
		\int
			\frac{
				1_{B_{1,3}^\complement}(\mathbf{k}_{n})
			}{
				|\mathbf{k}_{n}|^5
			}
			\,d\mathbf{k}_n
		\int
			\frac{
				1
			}{
				(\mathbf{k}_{j}^2 + 1)^4 |\mathbf{k}_{j}|
			}
			\,d\mathbf{k}_j.
	\label{chi_n_j_norm_bound_v1}
\end{aligned}
\end{equation}
These two integrals on the right-hand side can be evaluated using polar coordinates, giving us
\begin{equation}
\begin{aligned}
	\int
		\frac{
			1_{B_{1,3}^\complement}(\mathbf{k})
		}{
			|\mathbf{k}|^5
		}
		\,d\mathbf{k}
	=
		\int_1^\infty
			\frac{
				4\pi
			}{
				k^3
			}
			\,dk
	=
		\Big[
			\frac{
				-2\pi
			}{
				k^2
			}
		\Big]_{1}^{\infty}
	=
		2\pi,
	\label{k_5_integral}
\end{aligned}
\end{equation}
\begin{equation}
\begin{aligned}
	\int
		\frac{
			1
		}{
			(\mathbf{k}^2 + 1)^4
			|\mathbf{k}|
		}
		\,d\mathbf{k}
	=
		\int_0^\infty
			\frac{
				4\pi k
			}{
				(k^2 + 1)^4
			}
			\,dk
	=
		\Big[
			\frac{
				-2\pi
			}{
				3(k^2 + 1)^3
			}
		\Big]_{0}^{\infty}
	=
		\frac{
			2\pi
		}{
			3
		}
	<
		2\pi.
	\label{k_4_1_integral}
\end{aligned}
\end{equation}
And we thus obtain a bound on each $\|\chi_{n,j}\|^2$ given by
\begin{equation}
\begin{aligned}
	\|\chi_{n,j}\|^2 \leq
		\frac{(2\pi)^2 e^{-2n}}{n}
		\|\chi_{n-2}\|^2.
	\label{chi_n_j_norm_bound_rec}
\end{aligned}
\end{equation}
To then get a bound on $\|\chi_n\|^2 = \|\sum_{j=1}^{n-1}\chi_{n,j}\|^2$, we can use the fact that $\| v_1 + \ldots + v_n \|^2 \leq n^2 \max(\|v_1\|^2, \ldots, \|v_n\|^2)$ for any vectors $v_1, \ldots, v_n$ of a Hilbert space.\footnote{
	One can see this by noting that $\| v_1 + \ldots + v_n \|^2 = \sum_{i,j} \braket{v_i | v_ j} \leq \sum_{i,j} (\|v_i\|\|v_j\|) \leq n^2 \max_i(\|v_i\|^2)$.
}
Since $\chi_{n,j} \in L^2(\mathbb{R}^{3n+3})$, where we take $L^2(\mathbb{R}^{N})$ to denote the Hilbert space of complex-valued, square-integrable functions over $\mathbb{R}^{N}$ for any $N$, we therefore get that
\begin{equation}
\begin{aligned}
	\|\chi_n\|^2 \leq
		(n-1)^2 \max_{j\leq n-1}\big(\|\chi_{n,j}\|^2 \big)
	\leq
		\frac{(2\pi)^2 (n-1)^2 e^{-2n}}{n}
		\|\chi_{n-2}\|^2
	\leq
		(2\pi)^2 n e^{-2n}
		\|\chi_{n-2}\|^2.
	\label{chi_n_norm_bound_rec}
\end{aligned}
\end{equation}
And from this recursive relation, we obtain that
\begin{equation}
\begin{aligned}
	\|\chi_n\|^2 \leq
		\prod_{j=2,4,6,\ldots}^{n-m}\Big(
			(2\pi)^2
			j
			e^{-2(m+j)}
		\Big)
		\|\chi_{m}\|^2,
	\label{chi_n_norm_bound}
\end{aligned}
\end{equation}
which can be seen to be finite due to our assumption that $\|\chi_{m}\|^2 < \infty$.

We can now go back to Eq.\ (\ref{A_l_chi_n_j_norm_squared_v1}) and look at what we get in the cases when $l=j$, first of all. We see that when we shift $\mathbf{p}$ by $-\mathbf{k}_n$ for the integral, we get 
\begin{equation}
\begin{aligned}
	\|\hat A^-_{n,j} \chi_{n,j}\|^2 &=
		\frac{e^{-2n}}{n}
		\int
			d\mathbf{k}_n\,
			\frac{
				1_{B_{1,3}^\complement}(\mathbf{k}_{n})
			}{
				|\mathbf{k}_{n}|^5
			}
		\int d\mathbf{k}_1 \cdots \widehat{d\mathbf{k}_{j}} \cdots d\mathbf{k}_{n-1}\,
			d\mathbf{p}\,
			\frac{
				1
			}{
				\prod_{i=1, i\neq j}^{n-1} (\mathbf{k}_{i}^2 + 1)^4
			}
	\\&\phantom{=}\quad\quad\quad
		\bigg|
			\int
				\frac{
					f_{n}(\mathbf{k}_1, \ldots, \mathbf{k}_{n})
				}{
					(\mathbf{k}_{j}^2 + 1)^2 |\mathbf{k}_{j}|
				}
				\chi_{n-2}(
					\mathbf{k}_1, \ldots, \widehat{\mathbf{k}_{j}}, \ldots, \mathbf{k}_{n-1};
					\mathbf{p}
				)
				\,d\mathbf{k}_j
		\bigg|^2
	\label{A_l_eq_j_chi_n_j_norm_squared_v1}
\end{aligned}
\end{equation}
in these cases.
Next we see that since $\chi_{n-2}$ is square-integrable, as we have just shown, it means that
$
\mathbf{k}_j \mapsto
	\chi_{n-2}(
		\mathbf{k}_1, \ldots, \widehat{\mathbf{k}_{j}}, \ldots, \mathbf{k}_{n-1};
		\mathbf{p}
	)
$
is also square-integrable almost everywhere. We furthermore see that
$\mathbf{k}_j \mapsto 1/((\mathbf{k}_{j}^2 + 1)^2 |\mathbf{k}_{j}|)$ is square-integrable as well, since we have that
\begin{equation}
\begin{aligned}
	\int
		\frac{
			\big|f_{n}(\mathbf{k}_1, \ldots, \mathbf{k}_{n})\big|^2
		}{
			(\mathbf{k}^2 + 1)^4
			|\mathbf{k}|^2
		}
		\,d\mathbf{k}
	&\leq
		\int_0^\infty
			\frac{
				4\pi
			}{
				(k^2 + 1)^4
			}
			\,dk
%
	=
		\int_0^1
			\frac{
				4\pi
			}{
				(k^2 + 1)^4
			}
			\,dk
		+
		\int_1^\infty
			\frac{
				4\pi
			}{
				(k^2 + 1)^4
			}
			\,dk
	\\&=
		\Big[
			\frac{
				4\pi k
			}{
				(k^2 + 1)^4
			}
		\Big]_{0}^{1}
		-
		\int_0^1
			\frac{
				-32\pi k^2
			}{
				(k^2 + 1)^5
			}
			\,dk
		+
		\int_1^\infty
			\frac{
				4\pi
			}{
				(k^2 + 1)^4
			}
			\,dk
	\\&\leq
		\frac{\pi}{4}
		-
		\int_0^1
			\frac{
				-32\pi k
			}{
				(k^2 + 1)^5
			}
			\,dk
		+
		\int_1^\infty
			\frac{
				4\pi k
			}{
				(k^2 + 1)^4
			}
			\,dk
	\\&=
		\frac{\pi}{4}
		-
		\Big[
			\frac{
				4\pi
			}{
				(k^2 + 1)^4
			}
		\Big]_{0}^{1}
		+
		\Big[
			\frac{
				-2\pi
			}{
				3(k^2 + 1)^3
			}
		\Big]_{1}^{\infty}
	=
		4\pi +
		\frac{\pi}{12}
	<
		5\pi,
	\label{k_4_2_integral}
\end{aligned}
\end{equation}
where we have used the fact that $f_{n}(\mathbf{k}_1, \ldots, \mathbf{k}_{n}) \leq 1$ everywhere to get the first inequality, and also used integration by parts to get the second line. And since the inner product between two square-integrable functions is less than or equal to the product of their norms, we get that
\begin{equation}
\begin{aligned}
	\bigg|\int
		\frac{
			\chi_{n-2}(
				\mathbf{k}_1, \ldots, \widehat{\mathbf{k}_{j}}, \ldots, \mathbf{k}_{n-1};
				\mathbf{p}
			)
		}{
			f_{n}(\mathbf{k}_1, \ldots, \mathbf{k}_{n})^{-1}
			(\mathbf{k}_{j}^2 + 1)^2 |\mathbf{k}_{j}|
		}
		\,d\mathbf{k}_j
	\bigg|^2
	\leq
		5\pi \int d\mathbf{k}_j\,
			\big|\chi_{n-2}(
				\mathbf{k}_1, \ldots, \widehat{\mathbf{k}_{j}}, \ldots, \mathbf{k}_{n-1};
				\mathbf{p}
			)\big|^2
\end{aligned}
\end{equation}
almost everywhere.
Thus, we can see that we obtain the following bound for $\|\hat A^-_{n,j} \chi_{n,j}\|^2$:
\begin{equation}
\begin{aligned}
	\|\hat A^-_{n,j} \chi_{n,j}\|^2 &\leq
		\frac{5\pi e^{-2n}}{n}
		\|\chi_{n-2}\|^2
		\int
			\frac{
				1_{B_{1,3}^\complement}(\mathbf{k}_{n})
			}{
				|\mathbf{k}_{n}|^5
			}
			\,d\mathbf{k}_n
	\\&\phantom{=}\quad\quad\quad
		\int d\mathbf{k}_1 \cdots \widehat{d\mathbf{k}_{j}} \cdots d\mathbf{k}_{n-1}\,
			d\mathbf{p}\,
			\frac{
				\big|\chi_{n-2}(
					\mathbf{k}_1, \ldots, \widehat{\mathbf{k}_{j}}, \ldots, \mathbf{k}_{n-1};
					\mathbf{p}
				)\big|^2
			}{
				\prod_{i=1, i\neq j}^{n-1} (\mathbf{k}_{i}^2 + 1)^4
			}.
	\label{A_l_eq_j_chi_n_j_norm_bound_v1}
\end{aligned}
\end{equation}
We can then once again use the fact that the numerator in the last integrand here is real and non-negative to remove the denominator, since this will only increase the integral. When we then integrate over $\mathbf{k}_1, \ldots, \widehat{\mathbf{k}_{j}}, \ldots, \mathbf{k}_{n-1},\mathbf{p}$, we will thus get a factor of $\|\chi_{n-2}\|^2$ for our bound. And we can also use Eq.\ (\ref{k_5_integral}) once again, by which we get that
\begin{equation}
\begin{aligned}
	\|\hat A^-_{n,j} \chi_{n,j}\|^2 &\leq
		\frac{10\pi^2 e^{-2n}}{n}
		\|\chi_{n-2}\|^2.
	\label{A_l_eq_j_chi_n_j_norm_bound_v2}
\end{aligned}
\end{equation}
We then see that this bound is similar to the right-hand side of Eq.\ (\ref{chi_n_j_norm_bound_rec}). So by the same argument that gave us Eq.\ (\ref{chi_n_norm_bound}), we thus get that

\begin{equation}
\begin{aligned}
	\Big\|\sum_{j=1}^{n-1}\hat A^-_{n,j} \chi_{n,j}\Big\|^2 &\leq
		\prod_{j=2,4,6,\ldots}^{n-m}\Big(
			10\pi^2
			j
			e^{-2(m+j)}
		\Big)
		\|\chi_{m}\|^2.
	\label{sum_A_l_eq_j_chi_n_j_norm_bound}
\end{aligned}
\end{equation}

The remaining cases that are left to consider are the ones where $l\neq j$, $l < n$. 
From Eq.\ (\ref{A_l_chi_n_j_norm_squared_v1}), we see that we get
\begin{equation}
\begin{aligned}
	\|\hat A^-_{n,l} \chi_{n,j}\|^2 &=
		\frac{e^{-2n}}{n}
		\int
			d\mathbf{k}_n\,
			\frac{
				1_{B_{1,3}^\complement}(\mathbf{k}_{n})
			}{
				|\mathbf{k}_{n}|^5
			}
		\int d\mathbf{k}_1 \cdots \widehat{d\mathbf{k}_{l\,}} \cdots d\mathbf{k}_{n-1}\,
			d\mathbf{p}\,
			\frac{
				1
			}{
				\prod_{i=1, i\neq j, i\neq l}^{n-1} (\mathbf{k}_{i}^2 + 1)^4
			}
	\\&\phantom{=}\quad\quad\quad
			\frac{
				1
			}{
				(\mathbf{k}_{j}^2 + 1)^4 |\mathbf{k}_{j}|
			}
		\bigg|
			\int
				\frac{
					f_{n}(\mathbf{k}_1, \ldots, \mathbf{k}_{n})
				}{
					(\mathbf{k}_{l}^2 + 1)^2 \sqrt{|\mathbf{k}_{l}|}
				}
				\chi_{n-2}(
					\mathbf{k}_1, \ldots, \widehat{\mathbf{k}_{j}}, \ldots, \mathbf{k}_{n-1};
					\mathbf{p} - \mathbf{k}_l
				)
				\,d\mathbf{k}_l
		\bigg|^2
	\label{A_l_neq_j_chi_n_j_norm_squared_v1}
\end{aligned}
\end{equation}
in these cases, where we have also shifted $\mathbf{p}$ by $-\mathbf{k}_j - \mathbf{k}_n$ for the integral to get this expression.
We have already shown in Eq.\ (\ref{k_4_1_integral}) that 
$\mathbf{k}_j \mapsto 1/((\mathbf{k}_{j}^2 + 1)^2 \sqrt{|\mathbf{k}_{j}|})$ is a square-integrable function, namely with a squared norm less than $2\pi$.
We can therefore also use a similar argument here as the one that got us from Eq.\ (\ref{A_l_eq_j_chi_n_j_norm_squared_v1}) to Eq.\ (\ref{A_l_eq_j_chi_n_j_norm_bound_v1}), which will in this case give us that
\begin{equation}
\begin{aligned}
	\|\hat A^-_{n,l} \chi_{n,j}\|^2 &\leq
		\frac{2\pi e^{-2n}}{n}
		\|\chi_{n-2}\|^2
		\int
			\frac{
				1_{B_{1,3}^\complement}(\mathbf{k}_{n})
			}{
				|\mathbf{k}_{n}|^5
			}
			\,d\mathbf{k}_n
		\int
			\frac{
				1
			}{
				(\mathbf{k}_{j}^2 + 1)^4 |\mathbf{k}_{j}|
			}
			\,d\mathbf{k}_j
	\\&\phantom{=}\quad\quad\quad
		\int d\mathbf{k}_1 \cdots \widehat{d\mathbf{k}_{j}} \cdots d\mathbf{k}_{n-1}\,
			d\mathbf{p}\,
			\frac{
				\big|\chi_{n-2}(
					\mathbf{k}_1, \ldots, \widehat{\mathbf{k}_{j}}, \ldots, \mathbf{k}_{n-1};
					\mathbf{p} - \mathbf{k}_l
				)\big|^2
			}{
				\prod_{i=1, i\neq j, i \neq l}^{n-1} (\mathbf{k}_{i}^2 + 1)^4
			}.
	\label{A_l_neq_j_chi_n_j_norm_bound_v1}
\end{aligned}
\end{equation}
We can now shift $\mathbf{p}$ further by $\mathbf{k}_l$, use Eq.\ (\ref{k_5_integral}) and Eq.\ (\ref{k_4_1_integral}) once more, and also yet again use the fact that we are free to remove the denominator in the last integrand for our bound on $\|\hat A^-_{n,l} \chi_{n,j}\|^2$.
This yields us
\begin{equation}
\begin{aligned}
	\|\hat A^-_{n,l} \chi_{n,j}\|^2 \leq
		\frac{(2\pi)^3 e^{-2n}}{n}
		\|\chi_{n-2}\|^2,
	\quad
		l < n,\, l \neq j.
	\label{A_l_neq_j_chi_n_j_norm_bound_v2}
\end{aligned}
\end{equation}
And when we sum over all these cases, of which there are $(n-1)(n-2)$, we therefore get a combined bound that is $(n-1)^2(n-2)^2$ times greater than this bound. We thus get that
\begin{equation}
\begin{aligned}
	\Big\| \sum_{j=1}^{n-1} \sum_{l=1, l\neq j}^{n-1} \hat A^-_{n,l} \chi_{n,j} \Big\|^2 &\leq
		\frac{(2\pi)^3 (n-1)^2(n-2)^2 e^{-2n}}{n}
		\|\chi_{n-2}\|^2
	\\&\leq
		(2\pi)^3 n^3 e^{-2n}
		\prod_{j=2,4,6,\ldots}^{n-m-2}\Big(
			(2\pi)^2
			j
			e^{-2(m+j)}
		\Big)
		\|\chi_{m}\|^2
	\\&=
		2\pi n^2
		\prod_{j=2,4,6,\ldots}^{n-m}\Big(
			(2\pi)^2
			j
			e^{-2(m+j)}
		\Big)
		\|\chi_{m}\|^2,
	\label{sum_A_l_neq_j_chi_n_j_norm_bound}
\end{aligned}
\end{equation}
where we have also used Eq.\ (\ref{chi_n_norm_bound}) to get the second line.

We can now combine the bounds of Eqs.\ (\ref{sum_A_l_eq_j_chi_n_j_norm_bound}) and (\ref{sum_A_l_neq_j_chi_n_j_norm_bound}), yet again by using the fact that
$\| \sum_{i=1}^n v_i \|^2 \leq n^2 \max_i(\|v_i\|^2)$
for any vectors $v_1, \ldots, v_n$. This gives us that
\begin{equation}
\begin{aligned}
	\Big\| \sum_{l=1}^{n-1} \hat A^-_{n,l} \chi_{n} \Big\|^2 &\leq
		2^2\max\bigg(
			\Big\|\sum_{j=1}^{n-1}\hat A^-_{n,j} \chi_{n,j}\Big\|^2,\,
			\Big\| \sum_{j=1}^{n-1} \sum_{l=1, l\neq j}^{n-1} \hat A^-_{n,l} \chi_{n,j} \Big\|^2
		\bigg)
	\\&\leq
		8\pi n^2
		\prod_{j=2,4,6,\ldots}^{n-m}\Big(
			10\pi^2
			j
			e^{-2(m+j)}
		\Big)
		\|\chi_{m}\|^2.
	\label{A_l_less_than_n_chi_n_j_norm_bound}
\end{aligned}
\end{equation}
And it is easy to show that when we sum the right-hand side over all $n\in\{m+2, m+4, \ldots\}$, we will get $\|\chi_{m}\|^2$ times a finite positive constant. 
Furthermore, we can see that this sum will be maximal when $m=0$. This therefore tells us that there exists a positive constant, $C_1$, such that for all $m\in \mathbb{N}_0$ and $\ket{\chi} = \ket{\chi_m} + \ket{\chi_{m+2}} 
+ \ldots \in W$, we have
\begin{equation}
\begin{aligned}
	\Big\| \sum_{n=m+2}^{\infty} \sum_{l=1}^{n-1} \hat A^-_{n,l} \chi_{n} \Big\|^2 \leq
		C_1 \|\chi_{m}\|^2.
	\label{A_l_less_than_n_chi_n_j_norm_bound_simult}
\end{aligned}
\end{equation}
We will use this result below in Section \ref{sect_SA} as well.

To then conclude that $\|\hat A \chi\| < \infty$, all that is left to do is to see that we can write $\ket{\hat A \chi}$ as a sum of three vectors, namely as
\begin{equation}
\begin{aligned}
	\ket{\hat A \chi} =
		\ket{\hat A^-_m \chi_m} +
		\sum_{n=m+2}^{\infty} \ket{\hat A^-_{n,n} \chi_{n} + \hat A^+_{n-2} \chi_{n-2}} +
		\sum_{n=m+2}^{\infty} \sum_{l=1}^{n-1} \ket{\hat A^-_{n,l} \chi_{n}},
	\label{hat_A_chi_v1}
\end{aligned}
\end{equation}
where we have now shown that the last two of these vectors have finite norms. And we already have $\|\hat A^-_m \chi_m\| < \infty$ by assumption. It follows that $\|\hat A \chi\|$ is also finite, and since this applies for all $\ket{\chi} \in W$, we thus get that $W \subset U$.

\section[The density of the domain]{The density of $\mathrm{Dom}(\hat H_\mathrm{I})$ in $\mathbf{H}$} \label{sect_density}

In this section, we want to argue that the vectors in $W$ span a dense subspace of $\mathbf{H}$, which means that we can get arbitrarily close to any vector in $\mathbf{H}$, i.e.\ in the norm topology of $\mathbf{H}$, by taking the sum of some finite set of vectors in $W$. And since $W \subset \mathrm{Dom}(\hat H_\mathrm{I})$, this sum will thus also be a member of $\mathrm{Dom}(\hat H_\mathrm{I})$, giving us that $\mathrm{Dom}(\hat H_\mathrm{I})$ is dense in $\mathbf{H}$.

To show that $\mathrm{span}(W)$ is dense in $\mathbf{H}$, let us consider all $\ket{\chi} = \ket{\chi_m} + \ket{\chi_{m+2}} 
+ \ldots \in W$ where $\chi_m$ is smooth and has compact support bounded by some $R > 0$. Note that the set of all such $\chi_m$ functions spans $L^2(\mathbb{R}^{3m+3})$. For any such $\chi_m$, it is then easy to see that for any $(\mathbf{k}_1, \ldots, \mathbf{k}_{m+1}; \mathbf{p}) \in \mathrm{supp}(1_{F_{m+1} \times \mathbb{R}^3}\hat A^+_m \chi_{m})$,  $|\mathbf{k}_1|, \ldots, |\mathbf{k}_{m+1}|$ will all be bounded from above by $\exp[p_{m+1}(R,\ldots,R)] = \exp[\exp(m+1) \prod_{i=1}^{m} (R^2 + 1)^2]$. 
This means that there exists a sufficiently large $R'$ such that $F_{m+1} \subset B_{R',3}^{m+1}$. And we can therefore let $D_{m+1} \supset F_{m+1}$ be equal to $B_{R',3}^{m+1}$.

Let us then first confirm that such a $\ket{\chi}$ is a member of $W$. We can first of all note that the fact that $\chi_m$ is smooth and with compact support means that both $\|\chi_m\|$ and $\|\hat A^-_m \chi_m\|$ are finite. And since $D_{m+1} = B_{R',3}^{m+1}$ for some finite $R'$, we also get that $\| 1_{D_{m+1} \times \mathbb{R}^3}\hat A^+_m \chi_{m} \| < \infty$. And the only remaining constraint on $W$ that we need to check is then that $\ket{\chi} \in \mathbf{H}$ in the first place, which requires that $\|\chi\| < \infty$. But from Eq.\ (\ref{chi_n_norm_bound}), we see that
\begin{equation}
\begin{aligned}
	\|\chi\|^2 \leq
		\sum_{n\in\{m, m+2,m+4\ldots\}}^\infty
		\prod_{j=2,4,6,\ldots}^{n-m}\Big(
			(2\pi)^2
			j
			e^{-2(m+j)}
		\Big)
		\|\chi_{m}\|^2.
	\label{chi_norm_bound}
\end{aligned}
\end{equation}
And it is easy to show that this expression is finite whenever $\|\chi_m\|$ is. And in fact, we can also see, by a similar argument as we used to write up Eq.\ (\ref{A_l_less_than_n_chi_n_j_norm_bound_simult}), that there exists a positive constant, $C_2$, such that 
\begin{equation}
\begin{aligned}
	\|\chi\|^2 \leq
		C_2 \|\chi_{m}\|^2
	\label{chi_norm_bound_simult}
\end{aligned}
\end{equation}
for all $m\in \mathbb{N}_0$ and $\ket{\chi} = \ket{\chi_m} + \ket{\chi_{m+2}} 
+ \ldots \in W$.

If we then look at Eq.\ (\ref{chi_n}), we can see that $\chi_{n}(\mathbf{k}_1, \ldots, \mathbf{k}_{n}; \mathbf{p})$ for all $n>m$ is a sum of
$(n-1)(n-3)\cdots(m+1)$ terms.
And we see that all these include a factor of $1_{D_{m+1}^\complement}(\mathbf{k}_1', \ldots, \mathbf{k}_{m+1}')$, where $(\mathbf{k}_1', \ldots, \mathbf{k}_{n-1}')$ is some permutation of $(\mathbf{k}_1, \ldots, \mathbf{k}_{n-1})$ that differs for each term. Since we have that $D_{m+1}^\complement = (B_{R',3}^{m+1})^\complement$, if we therefore let $R'$ tend towards infinity, which we are allowed to do, we then see that each of these terms will tend to zero pointwise. Moreover, if we regard each term as the formula for a function in $L^2(\mathbb{R}^{3n+3})$, we see that all these functions will tend towards a zero vector in the norm topology of $L^2(\mathbb{R}^{3n+3})$. And we thus get that $\ket{\chi} \to \ket{\chi_m}$ for all such vectors when $R' \to \infty$.

Now, since the set of all such $\chi_m$ functions spans $L^2(\mathbb{R}^{3m+3})$, we can therefore get arbitrarily close to any particular $\ket{\psi_m}$ this way, $\psi \in \mathbf{H}'$. And since we can do this for any $m\in\mathbb{N}_0$, we thus get that the vectors in $W$ span a dense subspace of $\mathbf{H}$. Therefore, since $W \subset \mathrm{Dom}(\hat H_\mathrm{I})$, it follows that $\mathrm{Dom}(\hat H_\mathrm{I})$ is dense in $\mathbf{H}$.

\section[The symmetry of the operator]{The symmetry of $\hat H_\mathrm{I}$} \label{sect_sym}

In this section, we will show that the operator $\hat H_\mathrm{I}$, which we recall is given by $\hat H_\mathrm{I} \ket{\psi} = \ket{\hat A \psi}$ for all $\ket{\psi} \in \mathrm{Dom}(\hat H_\mathrm{I})$, is \emph{symmetric} in the sense that
\begin{equation}
\begin{aligned}
	 \braket{\hat A \phi | \psi} = \braket{\phi | \hat A \psi}
\end{aligned}
\end{equation}
for all $\ket{\phi}\!, \ket{\psi} \in \mathrm{Dom}(\hat H_\mathrm{I})$.


Let us recall, first of all, that
\begin{equation}
\begin{aligned}
	\hat A \psi = (
		\hat A^-_1 \psi_1,\;
		\hat A^+_0 \psi_0 + \hat A^-_2 \psi_2,\;
		\hat A^+_1 \psi_1 + \hat A^-_3 \psi_3,\,
		\ldots
	)
\end{aligned}
\end{equation}
with
\begin{equation}
\begin{aligned}
	\hat A^+_{n-1} \psi_{n-1}(\mathbf{k}_1, \ldots, \mathbf{k}_{n}; \mathbf{p}) =
		\frac{1}{\sqrt{n}}\sum_{i=1}^{n}
			\frac{1}{\sqrt{|\mathbf{k}_{i}|}}
			\psi_{n-1}(
				\mathbf{k}_1, \ldots, \widehat{\mathbf{k}_{i}}, \ldots, 
				\mathbf{k}_{n}; \mathbf{p} + \mathbf{k}_{i}
			),
\end{aligned}
\end{equation}
\begin{equation}
\begin{aligned}
	\hat A^-_{n} \psi_{n}(\mathbf{k}_1, \ldots, \mathbf{k}_{n-1}; \mathbf{p}) &=
	\frac{1}{\sqrt{n}}\sum_{i=1}^{n} \int
		\frac{1}{\sqrt{|\mathbf{k}_{n}|}}
		\psi_{n}(
			\mathbf{k}_1, \ldots, \mathbf{k}_{i-1}, \mathbf{k}_{n}, \mathbf{k}_{i}, \ldots,
			\mathbf{k}_{n-1}; \mathbf{p} - \mathbf{k}_n
		)
		\,d\mathbf{k}_{n}.
\end{aligned}
\end{equation}
To show that $\hat H_\mathrm{I}$ is symmetric, we thus want to show that the difference between the two series given by
\begin{equation}
\begin{aligned}
	\braket{\hat A \phi | \psi} &=
		\braket{\hat A^-_1 \phi_1 | \psi_0} +
		\braket{\hat A^+_0 \phi_0 + \hat A^-_2 \phi_2 | \psi_1} +
		\braket{\hat A^+_1 \phi_1 + \hat A^-_3 \phi_3 | \psi_2} +
		\ldots,
	\\
	\braket{\phi | \hat A \psi} &=
		\braket{\phi_0 | \hat A^-_1 \psi_1} +
		\braket{\phi_1 | \hat A^+_0 \psi_0 + \hat A^-_2 \psi_2} +
		\braket{\phi_2 | \hat A^+_1 \psi_1 + \hat A^-_3 \psi_3} +
		\ldots
\end{aligned}
\end{equation}
is zero.

Let us then consider the following cut-off versions of $\hat A^+_{n}$ and $\hat A^-_{n}$, call them $\hat A^{R+}_{n}$ and $\hat A^{R-}_{n}$, defined for all $\psi \in \mathbf{H}'$ and $n\in\mathbb{N}_+$ by
\begin{equation}
\begin{aligned}
	\hat A^{R+}_{n-1} \psi_{n-1} &=
		1_{(B_{R,3} \setminus B_{1/R,3})^{n} \times \mathbb{R}^3} \hat A^{+}_{n-1} \psi_{n-1},
	\\
	\hat A^{R-}_{n} \psi_n &=
		1_{(B_{R,3} \setminus B_{1/R,3})^{n-1} \times \mathbb{R}^3} \hat A^{-}_{n}
			(1_{(B_{R,3} \setminus B_{1/R,3})^{n} \times \mathbb{R}^3} \psi_n).
	\label{cut_off_hat_As}
\end{aligned}
\end{equation}
We see that this gives us
\begin{equation}
\begin{aligned}
	\braket{\hat A^{R+}_{n-1} \phi_{n-1} | \psi_{n}} &=
		\frac{1}{\sqrt{n}}\sum_{i=1}^{n}
		\int_{(B_{R,3} \setminus B_{1/R,3})^{n} \times \mathbb{R}^3}
			d\mathbf{k}_1 \cdots d\mathbf{k}_n\,d\mathbf{p}\,
			\frac{1}{\sqrt{|\mathbf{k}_{i}|}}
	\\&\phantom{=}\quad\quad\quad
		\phi_{n-1}(
			\mathbf{k}_1, \ldots, \widehat{\mathbf{k}_i}, \ldots, \mathbf{k}_{n};
			\mathbf{p} + \mathbf{k}_i
		)^* 
		\psi_{n}(\mathbf{k}_1, \ldots, \mathbf{k}_{n}; \mathbf{p}),
	\\
	\braket{\phi_{n-1} | \hat A^{R-}_{n} \psi_{n}} &=
		\frac{1}{\sqrt{n}}\sum_{i=1}^{n}
		\int_{(B_{R,3} \setminus B_{1/R,3})^{n} \times \mathbb{R}^3}
			d\mathbf{k}_1 \cdots d\mathbf{k}_n\,d\mathbf{p}\,
			\frac{1}{\sqrt{|\mathbf{k}_{n}|}}
	\\&\phantom{=}\quad\quad\quad
		\phi_{n-1}(\mathbf{k}_1, \ldots, \mathbf{k}_{n-1}; \mathbf{p})^* 
		\psi_{n}(
			\mathbf{k}_1, \ldots, \mathbf{k}_{i-1}, \mathbf{k}_{n}, \mathbf{k}_{i}, \ldots,
			\mathbf{k}_{n-1}; \mathbf{p} - \mathbf{k}_i
		).
\end{aligned}
\end{equation}
And from a simple renaming of $\mathbf{k}_1, \ldots, \mathbf{k}_{n}$,
we can easily see that these two expression are equal, giving us
\begin{equation}
\begin{aligned}
	\braket{\hat A^{R+}_{n-1} \phi_{n-1} | \psi_{n}} =
		\braket{\phi_{n-1} | \hat A^{R-}_{n} \psi_{n}}
	\label{cut_off_symmetry}
\end{aligned}
\end{equation}
for all $\phi, \psi \in \mathbf{H}'$, $n\in\mathbb{N}_+$, and $R \geq 1$.
Let us therefore define 
\begin{equation}
\begin{aligned}
	\braket{\hat A^R \phi | \psi}_{N} &=
		\sum_{n=0}^N \big(
			\braket{\hat A^{R+}_{n-1} \phi_{n-1} | \psi_{n}} +
			\braket{\hat A^{R-}_{n+1} \phi_{n+1} | \psi_{n}}
		\big),
	\\
	\braket{\phi | \hat A^R \psi}_{N} &=
		\sum_{n=0}^N \big(
			\braket{\phi_{n} | \hat A^{R+}_{n-1}  \psi_{n-1}} +
			\braket{\phi_{n} | \hat A^{R-}_{n+1} \psi_{n+1}}
		\big),
\end{aligned}
\end{equation}
where we let $\hat A^R$ be defined similarly as in Eq.\ (\ref{hat_A}), and let $\hat A^{R+}_{-1} = \hat A^{R-}_{0} = 0$. We then see that
\begin{equation}
\begin{aligned}
	\braket{\hat A^R \phi | \psi}_{N} - \braket{\phi | \hat A^R \psi}_{N} &=
		\braket{\hat A^{R-}_{N+1} \phi_{N+1} | \psi_{N}} +
		\braket{\phi_{N} | \hat A^{R-}_{N+1} \psi_{N+1}},
\end{aligned}
\end{equation}
namely since all other terms cancel each other. Thus, since the left-hand side of this equation converges to $\braket{\hat A \phi | \psi} - \braket{\phi | \hat A \psi}$ when $N, R \to \infty$, we get that
\begin{equation}
\begin{aligned}
	\braket{\hat A \phi | \psi} - \braket{\phi | \hat A \psi} =
		\lim_{R\to\infty} \lim_{N\to\infty} \big(
			\braket{\hat A^{R-}_{N+1} \phi_{N+1} | \psi_{N}} +
			\braket{\phi_{N} | \hat A^{R-}_{N+1} \psi_{N+1}}
		\big).
	\label{limit_of_symmetry_diviation}
\end{aligned}
\end{equation}
So if we can only show that $\braket{\phi_{N} | \hat A^{R-}_{N+1} \psi_{N+1}} 
\to 0$ when $N,R \to \infty$ for any $\ket{\phi}\!,\ket{\psi} \in \mathrm{Dom}(\hat H_\mathrm{I})$, we will get that both terms on the right-hand side of Eq.\ (\ref{limit_of_symmetry_diviation}) will tend to zero in this limit, which will thus give us that $\hat H_\mathrm{I}$ is symmetric.

In Section \ref{sect_W_subset_V}, we showed that
\begin{equation}
	F_{n-1} \times \mathbb{R}^3 \subset F_{n}^\complement
	\label{F_n_minus_1_productions_are_in_F_n_complement_v2}
\end{equation}
for all $n \geq 3$, which also implies that
\begin{equation}
	\mathcal{S} (F_{n-1} \times \mathbb{R}^3) \subset
		\mathcal{S} F_{n}^\complement
	=
		F_{n}^\complement.
	\label{F_n_minus_1_productions_are_in_F_n_complement_v3}
\end{equation}
And since $\hat A^+_N 1_{F_{N} \times \mathbb{R}^3} \phi_N$ is only supported on $\mathcal{S} (F_{N} \times \mathbb{R}^3)$, we thus get that $\hat A^+_N 1_{F_{N} \times \mathbb{R}^3} \phi_N$ is zero everywhere on $F_{N+1}$. So if we split $\phi$ and $\psi$ up into
\begin{equation}
\begin{aligned}
	\phi =   \hat 1_{T^\complement} \phi + \hat 1_{T} \phi
		\equiv \phi' + \phi'',
	\quad
	\psi =   \hat 1_{T^\complement} \psi + \hat 1_{T} \psi
		\equiv \psi' + \psi'',
	\label{phi_psi_primes}
\end{aligned}
\end{equation}
we therefore see that
\begin{equation}
\begin{aligned}
	\braket{\phi_{N}'' | \hat A^{R-}_{N+1} \psi_{N+1}''} =
		\braket{\hat A^{R+}_{N} \phi_{N}'' | \psi_{N+1}''}
	=
		0
\end{aligned}
\end{equation}
for all $N$ and $R$.

This motivates us to analyze $\braket{\hat A^{R+}_{N-1} \phi_{N-1} | \psi_{N}'}$, $\psi' = \hat 1_{T^\complement} \psi$, for all $\ket{\psi} \in U \cap V = \mathrm{Dom}(\hat H_\mathrm{I})$ and $\ket{\phi} \in \mathbf{H}$ in the hope to show that this inner product vanishes for all $R$ when $N \to \infty$.
In order to do this, let us first split up $\mathbb{R}^{3n}$ for all $n \in \mathbb{N}_+$ into a set of subsets, $\{X_{n,\mathbf i}\}_{\mathbf i \in \mathbb{N}_0^{n}}$, where $X_{n,\mathbf i} \subset \mathbb{R}^{3n}$ is given by
\begin{equation}
\begin{aligned}
	X_{n,\mathbf i} =
		(B_{i_1+1,3} \setminus B_{i_1,3}) \times \cdots \times
			(B_{i_n+1,3} \setminus B_{i_n,3})
\end{aligned}
\end{equation}
for all $\mathbf i = (i_1, \ldots, i_n) \in \mathbb{N}_0^n$.
Let us then split $\psi'_{N}$ up into $\psi'_{N} = \sum_{\mathbf i \in \mathbb{N}_0^{N}} \psi'_{N, \mathbf i}$, where
\begin{equation}
\begin{aligned}
	\psi'_{N, \mathbf i} = 1_{X_{N, \mathbf i} \times \mathbb{R}^3} \psi'_{N}.
\end{aligned}
\end{equation}
Below we will then show the following lemma: There exists a constant $C_3 > 0$ such that for any sufficiently large $N$, we have
\begin{equation}
	\| 1_{F_{N+1} \times \mathbb{R}^3} \hat A^+_{N} \psi'_{N} \|^2 \geq
		2\pi C_3 e^{ p'_{N+1}(i_1, \ldots, i_N) }
		\|\psi'_{N, \mathbf i}\|^2
	\label{hat_J_norm_lower_bound}
\end{equation}
for all $\mathbf i \in \mathbb{N}_0^{N}$, where we define $p'_n(k_1, \ldots, k_{n-1}) = p_n(k_1\hat{\mathbf{e}}_1, \ldots, k_{n-1}\hat{\mathbf{e}}_1)$, $\hat{\mathbf{e}}_1 = (1,0,0)$, in order to write this exponent more compactly.

Assuming that this lemma is true, let us see how this gives us that $\hat H_\mathrm{I}$ is symmetric.
First we note that
\begin{equation}
\begin{aligned}
	\braket{\hat A^{R+}_{N-1} \phi_{N-1} | \psi_{N, \mathbf i}'} =
		\braket{
			1_{X_{N, \mathbf i} \times \mathbb{R}^3} \hat A^{R+}_{N-1} \phi_{N-1} |
			\psi_{N, \mathbf i}'
		}
	\leq
		\|1_{X_{N, \mathbf i} \times \mathbb{R}^3} \hat A^{R+}_{N-1} \phi_{N-1}\|
		\|\psi_{N, \mathbf i}'\|.
	\label{braket_initial_bound}
\end{aligned}
\end{equation}
Now, because $\ket{\psi} \in V$, we know that $\sum_{n=0}^{\infty} \| 1_{F_{n+1} \times \mathbb{R}^3} \hat A^+_{n} \psi'_{n} \|^2$ converges, which means that for any $\varepsilon_1 > 0$,
we have
\begin{equation}
	\| 1_{F_{N+1} \times \mathbb{R}^3} \hat A^+_{N} \psi'_{N} \|^2 \leq \varepsilon_1
\end{equation}
for any sufficiently large $N$.
Combining this with Eq.\ (\ref{hat_J_norm_lower_bound}), we therefore get that
\begin{equation}
	\|\psi'_{N, \mathbf i}\|^2 \leq
		\frac{\varepsilon_1 }{2\pi C_3}
		e^{ -p'_{N+1}(i_1, \ldots, i_N) }
\end{equation}
for any $\mathbf i \in \mathbb{N}_0^{N}$

Let us then also derive an upper bound for $\|1_{X_{N, \mathbf i} \times \mathbb{R}^3} \hat A^{R+}_{N-1} \phi_{N-1}\|$. First let us define
\begin{equation}
\begin{aligned}
	\hat A^+_{n-1, l} \chi_{n-1}(\mathbf{k}_1, \ldots, \mathbf{k}_{n}; \mathbf{p}) =
		\frac{1}{\sqrt{n}}
			\frac{1}{\sqrt{|\mathbf{k}_{l}|}}
			\chi_{n-1}(
				\mathbf{k}_1, \ldots, \widehat{\mathbf{k}_{l\,}}, \ldots, 
				\mathbf{k}_{n}; \mathbf{p} + \mathbf{k}_{l}
			)
	\label{A_plus_n_l_minus_1}
\end{aligned}
\end{equation}
for all $\chi \in \mathbf{H}'$, $n\in\mathbb{N}_+$, $l\in\{1,\ldots, n\}$, such that $\hat A^+_{n-1} = \sum_{l=1}^{n} A^+_{n-1, l}$.
We then see that
\begin{equation}
\begin{aligned}
	\|1_{X_{N, \mathbf i} \times \mathbb{R}^3} \hat A^{R+}_{N-1,l} \phi_{N-1}\|^2 &=
		\frac{1}{N}
		\int_{
			X_{N, \mathbf i} \cap (B_{R,3} \setminus B_{1/R,3})^{N}
		} d\mathbf{k}_1 \cdots d\mathbf{k}_{N}\,
			\frac{1}{|\mathbf{k}_l|}
	\\&\phantom{=}\quad\quad\quad
		\int d\mathbf{p}\,
		\big|\phi_{N-1}(
			\mathbf{k}_1, \ldots, \widehat{\mathbf{k}_{l\,}}, \ldots, 
			\mathbf{k}_{N}; \mathbf{p}
		)\big|^2,
\end{aligned}
\end{equation}
where we have shifted $\mathbf{p}$ by $-\mathbf{k}_l$ for the last integral. Since the integrand is non-negative everywhere, we thus get that
\begin{equation}
\begin{aligned}
	\|1_{X_{N, \mathbf i} \times \mathbb{R}^3} \hat A^{R+}_{N-1,l} \phi_{N-1}\|^2 &\leq
		\frac{1}{N}
		\int_{
			i_l \leq |\mathbf{k}_l| \leq i_l + 1
		} d\mathbf{k}_1 \cdots d\mathbf{k}_{N}\,d\mathbf{p}\,
			\frac{1}{|\mathbf{k}_l|}
			\big|\phi_{N-1}(
				\mathbf{k}_1, \ldots, \widehat{\mathbf{k}_{l\,}}, \ldots, 
				\mathbf{k}_{N}; \mathbf{p}
			)\big|^2,
	\\&=
		\frac{1}{N} \|\phi_{N-1}\|^2
		\int_{i_l}^{i_l+1} 4\pi k \,dk
	=
		\frac{2\pi}{N}  \big((i_l + 1)^2 - i_l^2 \big) \|\phi_{N-1}\|^2
	\\&=
		\frac{2\pi}{N} (2i_l + 1) \|\phi_{N-1}\|^2.
\end{aligned}
\end{equation}
Furthermore, since $\phi \in \mathbf{H}'$, we know that $\sum_{n=0}^{\infty} \|\phi_n\|^2$ converges, which means that for all $\varepsilon_2 > 0$, 
we have $\|\phi_{N-1}\|^2 \leq \varepsilon_2$ for any sufficiently large $N$.
We therefore get that
\begin{equation}
\begin{aligned}
	\|1_{X_{N, \mathbf i} \times \mathbb{R}^3} \hat A^{R+}_{N-1,l} \phi_{N-1}\|^2 \leq
		\frac{2\pi}{N} (2i_l + 1) \varepsilon_2.
\end{aligned}
\end{equation}
And by using the fact again that $\| v_1 + \ldots + v_n \|^2 \leq n^2 \max(\|v_1\|^2, \ldots, \|v_n\|^2)$ for any vectors $v_1, \ldots, v_n$, we thus see that
\begin{equation}
\begin{aligned}
	\|1_{X_{N, \mathbf i} \times \mathbb{R}^3} \hat A^{R+}_{N-1} \phi_{N-1}\|^2 \leq
		2\pi N (2i_l + 1) \varepsilon_2.
\end{aligned}
\end{equation}
So from Eq.\ (\ref{braket_initial_bound}), we now get that
\begin{equation}
\begin{aligned}
	\braket{\hat A^{R+}_{N-1} \phi_{N-1} | \psi_{N, \mathbf i}'} \leq
		\sqrt{\frac{\varepsilon_1 \varepsilon_2}{C_3}}
		\sqrt{N(2i_l + 1)}
		e^{ -p'_{N+1}(i_1, \ldots, i_N)/2 }
	\label{braket_bound_v1}
\end{aligned}
\end{equation}
for all $\mathbf i \in \mathbb{N}_0^{N}$.
And it is not hard to see that
$
\sum_{\mathbf i \in \mathbb{N}_0^{N}} \sqrt{N(2i_l + 1)}
	\exp( -p'_{N+1}(i_1, \ldots, i_N)/2 )
<
	\infty
$
from the formula of Eq.\ (\ref{p_n}),
which then tells us that
\begin{equation}
\begin{aligned}
	\braket{\hat A^{R+}_{N-1} \phi_{N-1} | \psi_{N}'} =
		\sum_{\mathbf i \in \mathbb{N}_0^{N}}
			\braket{\hat A^{R+}_{N-1} \phi_{N-1} | \psi_{N, \mathbf i}'}
	\leq
		C_4 \sqrt{\varepsilon_1 \varepsilon_2}
\end{aligned}
\end{equation}
for some constant $C_4 \geq 0$. Thus, since we can choose arbitrarily small $\varepsilon_1$ and $\varepsilon_2$ for a large enough $N$, we get that
\begin{equation}
\begin{aligned}
	\lim_{N\to \infty} \braket{\hat A^{R+}_{N-1} \phi_{N-1} | \psi_{N}'} = 0
\end{aligned}
\end{equation}
for all $R \geq 1$, $\ket{\phi} \in \mathbf{H}$, $\ket{\psi} \in \mathrm{Dom}(\hat H_\mathrm{I})$, $\psi' = \hat 1_{T^\complement} \psi$.

From this result, we see that for any $\ket{\phi}\!, \ket{\psi} \in \mathrm{Dom}(\hat H_\mathrm{I})$, we have that $\braket{\hat A^{R+}_{N} \phi_{N}'' | \psi_{N+1}'}$, $\braket{\hat A^{R+}_{N} \phi_{N}' | \psi_{N+1}''}$, and $\braket{\hat A^{R+}_{N} \phi_{N}' | \psi_{N+1}'}$ all tend towards zero when $N \to \infty$, where $\phi', \phi'', \psi', \psi''$ are defined according to Eq.\ (\ref{phi_psi_primes}). And we have already shown that $\braket{\hat A^{R+}_{N} \phi_{N}'' | \psi_{N+1}''} = 0$. We therefore get that
\begin{equation}
\begin{aligned}
	\lim_{N\to \infty} \braket{\hat A^{R+}_{N-1} \phi_{N-1} | \psi_{N}} =
	\lim_{N\to \infty} \braket{\phi_{N-1} | \hat A^{R-}_{N} \psi_{N}} =
	0
\end{aligned}
\end{equation}
for all $\ket{\phi}\!, \ket{\psi} \in \mathrm{Dom}(\hat H_\mathrm{I})$. And this tells us that both terms on the right-hand side of Eq.\ (\ref{limit_of_symmetry_diviation}) vanishes in the limit when $N\to \infty$. This means that the left-hand side also vanishes, and we therefore get that
\begin{equation}
\begin{aligned}
	\braket{\hat A \phi | \psi} - \braket{\phi | \hat A \psi} = 0,
\end{aligned}
\end{equation}
which is exactly what we want to show in order to conclude that $\hat H_\mathrm{I}$ is symmetric.

So all that is left in order to show that $\hat H_\mathrm{I}$ is symmetric is to show the lemma of Eq.\ (\ref{hat_J_norm_lower_bound}), which stated that for all $\ket{\psi} \in U \cap V$ and $\ket{\phi} \in \mathbf{H}$, there exists a positive constant $C_3$ such that for any $N\in\mathbb{N}$ that is sufficiently large,
\begin{equation}
	\| 1_{F_{N+1} \times \mathbb{R}^3} \hat A^+_{N} \psi'_{N} \|^2 \geq
		2\pi C_3 e^{ p'_{N+1}(i_1, \ldots, i_N) }
		\|\psi'_{N, \mathbf i}\|^2
	\label{hat_J_norm_lower_bound_v2}
\end{equation}
for all $\mathbf i \in \mathbb{N}_0^{N}$, where $\psi' = \hat 1_{T^\complement} \psi$.

In order to do this, let us first define $F_{N,j}$ for all $N,j\in\mathbb{N}_+$, $j\leq N$, to be the set of all $(\mathbf{k}_1, \ldots, \mathbf{k}_{N}) \in F_N$ for which $\mathbf{k}_j$ has the greatest norm. Note that this is equivalent of requiring that $(\mathbf{k}_1, \ldots, \widehat{\mathbf{k}_{j}}, \ldots, \mathbf{k}_{n}, \mathbf{k}_{j}) \in E_n$.
Let us then also define $E'_{N, j}$ to be the set given by $E'_{N, j} = \{(\mathbf{k}_1, \ldots, \mathbf{k}_{N}) \,|\,(\mathbf{k}_1, \ldots, \widehat{\mathbf{k}_{j}}, \ldots, \mathbf{k}_{N}, \mathbf{k}_{j}) \in E_N\}$, such that $F_{N,j} = F_N \cap E'_{N, j}$ for all $N$ and $j\in \{1, \ldots, N\}$.
We then want to analyze the various contributions to $\| 1_{F_{N+1} \times \mathbb{R}^3} \hat A^+_{N} \psi'_{N} \|^2$ given by $\| 1_{F_{N+1, n} \times \mathbb{R}^3} \hat A^+_{N,l} \psi'_{N, \mathbf i} \|^2$ for $n,l \in \{1, \ldots, N+1\}$ and $\mathbf i \in \mathbb{N}_0^{N}$.

Let us start by analyzing the cases when $l = n$. Since $\psi'_{N, \mathbf i}$ is supported only on $F_N^\complement \times \mathbb{R}^3$, we first of all see that
\begin{equation}
\begin{aligned}
	\mathrm{supp}(\hat A^+_{N,n} \psi'_{N, \mathbf i}) \subset
		\big(
			\mathcal{P}_{n\leftrightarrow N+1} (F_N^\complement \times \mathbb{R}^3)
		\big)
			\times \mathbb{R}^3,
\end{aligned}
\end{equation}
where we define $\mathcal{P}_{i\leftrightarrow j}$ as an operator that permutes the $i$th coordinate vector with the $j$th coordinate vector for all elements.\footnote{
	In other words, $\mathcal{P}_{i\leftrightarrow j} X = \{(\mathbf x_1, \ldots, \mathbf x_n) \,|\, (\mathbf x_1, \ldots, \mathbf x_{i-1}, \mathbf x_{j}, \mathbf x_{i+1}, \ldots, \mathbf x_{j-1}, \mathbf x_{i}, \mathbf x_{j+1}, \ldots, \mathbf x_n) \in X \}$.
}
And since
\begin{equation}
\begin{aligned}
	F_{N+1,n} =
		\Big(\mathcal{S} \big( (F_{N}^\complement \times \mathbb{R}^3) \cap E_{N+1} \big)\Big)
		\cap
		E'_{N+1, n}
	=
		\big(\mathcal{P}_{n\leftrightarrow N+1} (F_N^\complement \times \mathbb{R}^3)\big)
		\cap
		E'_{N+1, n},
\end{aligned}
\end{equation}
we thus get that
\begin{equation}
\begin{aligned}
	1_{F_{N+1, n} \times \mathbb{R}^3} \hat A^+_{N,n} \psi'_{N, \mathbf i} =
		1_{E'_{N+1, n} \times \mathbb{R}^3} \hat A^+_{N,n} \psi'_{N, \mathbf i}.
\end{aligned}
\end{equation}
This then allows us to write
\begin{equation}
\begin{aligned}
	\| 1_{F_{N+1,n} \times \mathbb{R}^3} \hat A^+_{N,n} \psi'_{N, \mathbf i} \|^2 &=
		\frac{1}{N+1}
		\int_{X_{N, \mathbf i}}
			d\mathbf{k}_1 \cdots \widehat{d\mathbf{k}_{n}} \cdots d\mathbf{k}_{N+1}
		\int_{
			(\mathbf{k}_1, \ldots, \mathbf{k}_{N+1}) \in E'_{N+1, n}
		} d\mathbf{k}_n\,
			\frac{1}{|\mathbf{k}_n|}
	\\&\phantom{=}\quad\quad\quad
		\int d\mathbf{p}\,
		\big|\psi'_{N, \mathbf i}(
			\mathbf{k}_1, \ldots, \widehat{\mathbf{k}_{n}}, \ldots, 
			\mathbf{k}_{N+1}; \mathbf{p}
		)\big|^2.
\end{aligned}
\end{equation}
Here we have thus also shifted $\mathbf{p}$ by $-\mathbf{k}_n$ for the last integral, and we have also used the fact that $\psi'_{N, \mathbf i}$ is only supported on $X_{N, \mathbf i} \times \mathbb{R}^3$ to 
limit the first integral to $(\mathbf{k}_1, \ldots, \widehat{\mathbf{k}_{n}}, \ldots, \mathbf{k}_{N+1}) \in X_{N, \mathbf i}$.
Then for a sufficiently large $N$, we see that
\begin{equation}
\begin{aligned}
	\int_{
		(\mathbf{k}_1, \ldots, \mathbf{k}_{N+1}) \in E'_{N+1, n}
	}
		\frac{1}{|\mathbf{k}_n|}
		\,d\mathbf{k}_n
	&=
	\int_{
		e^{
			p_{N+1}(
				\mathbf{k}_1, \ldots, \widehat{\mathbf{k}_{n}}, \ldots, \mathbf{k}_{N+1}
			)/2
		}
	}^{
		e^{
			p_{N+1}(
				\mathbf{k}_1, \ldots, \widehat{\mathbf{k}_{n}}, \ldots, \mathbf{k}_{N+1}
			)
		}
	}
		4\pi k
		\,dk
	\\&=
		\Big[
			2\pi k^2
		\Big]_{
			e^{
				p_{N+1}(
					\mathbf{k}_1, \ldots, \widehat{\mathbf{k}_{n}}, \ldots, \mathbf{k}_{N+1}
				)/2
			}
		}^{
			e^{
				p_{N+1}(
					\mathbf{k}_1, \ldots, \widehat{\mathbf{k}_{n}}, \ldots, \mathbf{k}_{N+1}
				)
			}
		}
	\\&\geq
		2\pi
		e^{
			p_{N+1}(
				\mathbf{k}_1, \ldots, \widehat{\mathbf{k}_{n}}, \ldots, \mathbf{k}_{N+1}
			)
		},
\end{aligned}
\end{equation}
where we for the last inequality have used the fact that for a sufficiently large $x$, we have
$
	\exp(2x) - \exp(x) =
		\exp(x) (\exp(x) - 1)
	\geq
		\exp(x)
$.
We thus obtain that
\begin{equation}
\begin{aligned}
	\| 1_{F_{N+1,n} \times \mathbb{R}^3} \hat A^+_{N,n} \psi'_{N, \mathbf i} \|^2 &\geq
		\frac{2\pi}{N+1}
		\int_{X_{N, \mathbf i}}
			d\mathbf{k}_1 \cdots \widehat{d\mathbf{k}_{n}} \cdots d\mathbf{k}_{N+1}\,
		e^{
			p_{N+1}(
				\mathbf{k}_1, \ldots, \widehat{\mathbf{k}_{n}}, \ldots, \mathbf{k}_{N+1}
			)
		}
	\\&\phantom{=}\quad\quad\quad
		\big|\psi'_{N, \mathbf i}(
			\mathbf{k}_1, \ldots, \widehat{\mathbf{k}_{n}}, \ldots, 
			\mathbf{k}_{N+1}; \mathbf{p}
		)\big|^2
	\\&\geq
		\frac{2\pi}{N+1}
		e^{ p'_{N+1}(i_1, \ldots, i_N) }
		\|\psi'_{N, \mathbf i}\|^2.
	\label{F_n_A_n_lower_bound_v1}
\end{aligned}
\end{equation}

Next we want to show that
any two functions $1_{F_{N+1, n} \times \mathbb{R}^3} \hat A^+_{N,n} \psi'_{N, \mathbf i}$ and $1_{F_{N+1, m} \times \mathbb{R}^3} \hat A^+_{N,m} \psi'_{N,\, \mathbf j}$ will have no overlap in terms of their support when $\mathbf j \neq \mathbf i$ or $m \neq n$. First we see that for any $(\mathbf{k}_1, \ldots, \mathbf{k}_{N+1})$ in the support of $1_{F_{N+1, n} \times \mathbb{R}^3} \hat A^+_{N,n} \psi'_{N, \mathbf i}$, $|\mathbf{k}_n| = \max(|\mathbf{k}_1|, \ldots, |\mathbf{k}_{N+1}|)$, and we can say a similar thing for $|\mathbf{k}_m|$ in the support of $1_{F_{N+1, m} \times \mathbb{R}^3} \hat A^+_{N,m} \psi'_{N,\, \mathbf j}$. Therefore we cannot have $m \neq n$ if the functions are to overlap. And for $m = n$, we see that $(\mathbf{k}_1, \ldots, \widehat{\mathbf{k}_n}, \ldots, \mathbf{k}_{N+1})$ must be a member both $X_{N, \mathbf i}$ and $X_{N, \mathbf j}$ in order to be in support of both of these two functions at the same time, which is only possible when $\mathbf j = \mathbf i$.
This result then tells us that
\begin{equation}
\begin{aligned}
	\Big\|
		\sum_{\mathbf i \in \mathbb{N}_0^N}^{\mathbf{i}^2 < M} \sum_{n=1}^{N+1}
		1_{F_{N+1,n} \times \mathbb{R}^3} \hat A^+_{N,n} \psi'_{N, \mathbf i}
	\Big\|^2 &=
		\sum_{\mathbf i \in \mathbb{N}_0^N}^{\mathbf{i}^2 < M} \sum_{n=1}^{N+1}
		\| 1_{F_{N+1,n} \times \mathbb{R}^3} \hat A^+_{N,n} \psi'_{N, \mathbf i} \|^2
	\\&\geq
		\sum_{\mathbf i \in \mathbb{N}_0^N}^{\mathbf{i}^2 < M}
		2\pi e^{ p'_{N+1}(i_1, \ldots, i_N) }
		\|\psi'_{N, \mathbf i}\|^2
	\label{F_n_A_n_lower_bound_sum}
\end{aligned}
\end{equation}
for any $M\in\mathbb{N}$.

We have thus obtained a lower bound for the norm of the sum of all $1_{F_{N+1, n} \times \mathbb{R}^3} \hat A^+_{N,l} \psi'_{N, \mathbf i}$-functions with $l = n$. Note that this expression looks somewhat similar to the right-hand side of Eq.\ (\ref{hat_J_norm_lower_bound_v2}). 
Next we want to look for an upper bound on the norms of such functions when $l \neq n$. Our hope is to show that these are insignificant in comparison, by which we will obtain our lemma.

For $\| 1_{F_{N+1, n} \times \mathbb{R}^3} \hat A^+_{N,l} \psi'_{N, \mathbf i} \|^2$, $n \neq l$, we see that
\begin{equation}
\begin{aligned}
	\| 1_{F_{N+1,n} \times \mathbb{R}^3} \hat A^+_{N,l} \psi'_{N, \mathbf i} \|^2 &=
		\frac{1}{N+1}
		\int_{X_{N, \mathbf i}}
			d\mathbf{k}_1 \cdots \widehat{d\mathbf{k}_{l}} \cdots d\mathbf{k}_{N+1}
		\int_{
			(\mathbf{k}_1, \ldots, \mathbf{k}_{N+1}) \in E'_{N+1, n}
		} d\mathbf{k}_l\,
			\frac{1}{|\mathbf{k}_l|}
	\\&\phantom{=}\quad\quad\quad
		\int d\mathbf{p}\,
		\big|\psi'_{N, \mathbf i}(
			\mathbf{k}_1, \ldots, \widehat{\mathbf{k}_{l\,}}, \ldots, 
			\mathbf{k}_{N+1}; \mathbf{p}
		)\big|^2.
\end{aligned}
\end{equation}
%
And from Eqs. (\ref{p_n}--\ref{E_n}) we have that
\begin{equation}
\begin{aligned}
	\frac{1}{2} e^{n} \prod_{j=1}^{n-1} (\mathbf{k}_{j}^2 + 1)^2 <
		\ln|\mathbf{k}_n|
\end{aligned}
\end{equation}
for all $(\mathbf{k}_{1}, \ldots, \mathbf{k}_{n}) \in E_n$,
which implies that
\begin{equation}
\begin{aligned}
	\mathbf{k}_{l}^4 <
		(\mathbf{k}_{l}^2 + 1)^2
	<
		2e^{-n} \ln|\mathbf{k}_n|
		\prod_{j=1, j\neq l}^{n-1} (\mathbf{k}_{j}^2 + 1)^{-2}
\end{aligned}
\end{equation}
for any $l\in\{1,\ldots,n-1\}$.
Therefore, since $1/|\mathbf{k}_l|$ is positive everywhere, we get that
\begin{equation}
\begin{aligned}
	\int_{
		(\mathbf{k}_1, \ldots, \mathbf{k}_{N+1}) \in E'_{N+1, n}
	}
		\frac{1}{|\mathbf{k}_l|}
		\,d\mathbf{k}_l
	&\leq
		\int_{
			0
		}^{
			(
				2e^{-(N+1)} \ln|\mathbf{k}_n|
				\prod_{j=1, j\neq n, j\neq l}^{N+1} (\mathbf{k}_{j}^2 + 1)^{-2}
			)^{1/4}
		}
			4\pi k
			\,dk
	\\&=
		2\pi
		\big(
			2e^{-(N+1)} \ln|\mathbf{k}_n|
			\prod_{j=1, j\neq n, j\neq l}^{N+1} (\mathbf{k}_{j}^2 + 1)^{-2}
		\big)^{1/2}
	\\&=
		2\sqrt{2}\pi e^{-(N+1)/2}
		\sqrt{\ln|\mathbf{k}_n|}
		\prod_{j=1, j\neq n, j\neq l}^{N+1} (\mathbf{k}_{j}^2 + 1)^{-1}.
\end{aligned}
\end{equation}
And thus we obtain that
\begin{equation}
\begin{aligned}
	\| 1_{F_{N+1, n} \times \mathbb{R}^3} \hat A^+_{N,l} \psi'_{N, \mathbf i} \|^2 &\leq
		2\sqrt{2}\pi e^{-(N+1)/2}
		\int_{X_{N, \mathbf i}}
			d\mathbf{k}_1 \cdots \widehat{d\mathbf{k}_{l}} \cdots d\mathbf{k}_{N+1}\,
			d\mathbf{p}\,
	\\&\phantom{=}\quad\quad\quad
		\frac{\sqrt{\ln|\mathbf{k}_n|}}{
			\prod_{j=1, j\neq n, j\neq l}^{N+1} (\mathbf{k}_{j}^2 + 1)
		}
		\big|\psi'_{N, \mathbf i}(
			\mathbf{k}_1, \ldots, \widehat{\mathbf{k}_{l\,}}, \ldots, 
			\mathbf{k}_{N+1}; \mathbf{p}
		)\big|^2
	\\&\leq
		2\sqrt{2}\pi e^{-(N+1)/2}
		\frac{\sqrt{\ln(i_n + 1)}}{
			\prod_{j=1, j\neq n}^{N} (i_j^2 + 1)
		}
		\|\psi'_{N, \mathbf i}\|^2
	\\&=
		\frac{
			2\sqrt{2}\pi
			(i_n + 1)
			\sqrt{\ln(i_n + 1)}
		}{
			\sqrt{p'_{N+1}(i_1, \ldots, i_N)}
		}
		\|\psi'_{N, \mathbf i}\|^2
	\label{F_n_A_l_upper_bound_v1}
\end{aligned}
\end{equation}
for $l \neq n$.

This upper bound does indeed seem insignificant compared to the lower bound of Eq.\ (\ref{F_n_A_n_lower_bound_v1}). However, these $1_{F_{N+1, n} \times \mathbb{R}^3} \hat A^+_{N,l} \psi'_{N, \mathbf i}$-functions with $l \neq n$ might overlap, which might increase the norm of their sum. Fortunately, we can see that for any given coordinate, $(\mathbf{k}_1, \ldots, \mathbf{k}_{N+1})$, there are only $N$ values of $\mathbf{i}$ such that $1_{F_{N+1, n} \times \mathbb{R}^3} \hat A^+_{N,l} \psi'_{N, \mathbf i}$, $l \neq n$, can be supported in that coordinate. This is because we need to be able to pick out a $\mathbf{k}_l$, $l\neq n$, from $(\mathbf{k}_1, \ldots, \mathbf{k}_{N+1})$ and then have $(\mathbf{k}_1, \ldots, \widehat{\mathbf{k}_{l\,}}, \ldots, \mathbf{k}_{N+1}) \in X_{N, \mathbf i}$ in order for this to be possible. This means that the contribution to $\|\sum_{\mathbf i \in N_0^N}^{\mathbf{i}^2 < M} 1_{F_{N+1, n} \times \mathbb{R}^3} \hat A^+_{N,l} \psi'_{N, \mathbf i}\|^2$ from neighborhood of $(\mathbf{k}_1, \ldots, \mathbf{k}_{N+1})$ can be no more than $N^2$ times\footnote{
	Here we are thus again using the fact that $\| v_1 + \ldots + v_n \|^2 \leq n^2 \max(\|v_1\|^2, \ldots, \|v_n\|^2)$ for any vectors $v_1, \ldots, v_n$,
	which implies that $\| v_1 + \ldots + v_n \|^2 \leq n^2 (\|v_1\|^2 + \ldots + \|v_n\|^2)$.
	This bound is in fact not even optimal, as one can also show that $\| v_1 + \ldots + v_n \|^2 \leq n (\|v_1\|^2 + \ldots + \|v_n\|^2)$.
}
greater than its contribution to $\sum_{\mathbf i \in N_0^N}^{\mathbf{i}^2 < M} \|1_{F_{N+1, n} \times \mathbb{R}^3} \hat A^+_{N,l} \psi'_{N, \mathbf i}\|^2$.
And we thus get that
\begin{equation}
\begin{aligned}
	\Big\|
		\sum_{\mathbf i \in \mathbb{N}_0^N}^{\mathbf{i}^2 < M}
		\sum_{l=1}^{N+1} \sum_{n=1, n\neq l}^{N+1}
			1_{F_{N+1, n} \times \mathbb{R}^3} \hat A^+_{N,l} \psi'_{N, \mathbf i}
	\Big\|^2 &\leq
		N^2
		\sum_{\mathbf i \in \mathbb{N}_0^N}^{\mathbf{i}^2 < M}
		\Big\|
			\sum_{l=1}^{N+1} \sum_{n=1, n\neq l}^{N+1}
				1_{F_{N+1, n} \times \mathbb{R}^3} \hat A^+_{N,l} \psi'_{N, \mathbf i}
		\Big\|^2
	\\&\leq
		N^2 (N+1)^4
		\sum_{\mathbf i \in \mathbb{N}_0^N}^{\mathbf{i}^2 < M}
		\max_{n\neq l}\Big(
			\| 1_{F_{N+1, n} \times \mathbb{R}^3} \hat A^+_{N,l} \psi'_{N, \mathbf i} \|^2
		\Big)
	\\&\leq
		\sum_{\mathbf i \in \mathbb{N}_0^N}^{\mathbf{i}^2 < M}
		\frac{
			2\sqrt{2}\pi
			(N+1)^6
			(i_n + 1)
			\sqrt{\ln(i_n + 1)}
		}{
			\sqrt{p'_{N+1}(i_1, \ldots, i_N)}
		}
		\|\psi'_{N, \mathbf i}\|^2.
	\label{F_n_A_l_upper_bound_sum}
\end{aligned}
\end{equation}
We see that the right-hand side here is still much smaller, for sufficiently large $N$, than the right-hand side of Eq.\ (\ref{F_n_A_n_lower_bound_sum}). Indeed, it is not hard to show that 
for any $\varepsilon_3 < 1$, there is a sufficiently large $N$ such that
\begin{equation}
\begin{aligned}
	\frac{
		2\sqrt{2}\pi
		(N+1)^6
		(i_n + 1)
		\sqrt{\ln(i_n + 1)}
	}{
		\sqrt{p'_{N+1}(i_1, \ldots, i_N)}
	}
	\leq
		\varepsilon_3
		2\pi e^{ p'_{N+1}(i_1, \ldots, i_N) }
	\label{epsilon_3_bound}
\end{aligned}
\end{equation}
for all $\mathbf i \in \mathbb{N}_0^{N}$.
This then tells us, combining Eqs.\ (\ref{F_n_A_n_lower_bound_sum}), (\ref{F_n_A_l_upper_bound_sum}), and (\ref{epsilon_3_bound}), that
\begin{equation}
\begin{aligned}
	\Big\|
		\sum_{\mathbf i \in \mathbb{N}_0^N}^{\mathbf{i}^2 < M}
			1_{F_{N+1} \times \mathbb{R}^3} \hat A^+_{N} \psi'_{N, \mathbf i}
	\Big\|^2 &\geq
		(1-\sqrt{\varepsilon_3})^2
		\sum_{\mathbf i \in \mathbb{N}_0^N}^{\mathbf{i}^2 < M}
		2\pi e^{ p'_{N+1}(i_1, \ldots, i_N) }
		\|\psi'_{N, \mathbf i}\|^2.
\end{aligned}
\end{equation}
To obtain the lemma of Eq.\ (\ref{hat_J_norm_lower_bound_v2}), we then see that since $\ket{\psi} \in V$, the left-hand side must converge when $M\to \infty$, namely to $\| 1_{F_{N+1} \times \mathbb{R}^3} \hat A^+_{N} \psi'_{N} \|^2$. And we then also see that
\begin{equation}
\begin{aligned}
	\lim_{M\to\infty}
	\sum_{\mathbf j \in \mathbb{N}_0^N}^{\mathbf{j}^2 < M}
	2\pi e^{ p'_{N+1}(i_1, \ldots, i_N) }
	\|\psi'_{N, \mathbf j}\|^2
	\geq
		2\pi e^{ p'_{N+1}(i_1, \ldots, i_N) }
		\|\psi'_{N, \mathbf i}\|^2
\end{aligned}
\end{equation}
for any $\mathbf{i} \in \mathbb{N}_0^{N}$.
Thus, if we define $C_3 = (1-\sqrt{\varepsilon_3})^2$, we obtain that
\begin{equation}
\begin{aligned}
		\| 1_{F_{N+1} \times \mathbb{R}^3} \hat A^+_{N} \psi'_{N} \|^2
	\geq
		C_3
		2\pi e^{ p'_{N+1}(i_1, \ldots, i_N) }
		\|\psi'_{N, \mathbf i}\|^2
\end{aligned}
\end{equation}
for all $\mathbf i \in \mathbb{N}_0^{N}$, which is what we wanted to show.
This was the last piece that we needed, and we therefore get that $\hat H_\mathrm{I}$ is symmetric.

\section[The self-adjointness of the operator]{The self-adjointness of $\hat H_\mathrm{I}$} \label{sect_SA}

We now want to finally show that $\hat H_\mathrm{I}$ is self-adjoint on its domain, which will thus give us Proposition \ref{prop_1}.

We have the following definitions regarding self-adjointness according to Hall \cite{Hall}.
First of all, an \emph{unbounded operator}, $\hat T$, on $\mathbf{H}$ is any linear operator where $\mathrm{Dom}(\hat T)$ is a dense subspace of $\mathbf{H}$ and $\hat T \psi \in \mathbf{H}$ for all $\psi \in \mathrm{Dom}(\hat T)$. And the \emph{adjoint}, $\hat T^*$, of such an operator is defined as the operator for which $\mathrm{Dom}(\hat T^*)$ is equal to the subspace of all $\phi \in \mathbf{H}$ where
\begin{equation}
\begin{aligned}
	\psi \mapsto
		\frac{\braket{\phi | \hat T \psi}}{\|\psi\|},
	\quad
	\psi \in \mathrm{Dom}(\hat T),
	\quad
	\|\psi\| > 0,
	\label{functional_adjoint_domain}
\end{aligned}
\end{equation}
is bounded as a function,\footnote{
	Hall \cite{Hall} uses the terminology of a \emph{linear functional} with which the division by $\|\psi\|$ is implicit.
}
and where $\braket{\phi | \hat T \psi} = \braket{\hat T^* \phi | \psi}$ for all $\phi \in \mathrm{Dom}(\hat T^*)$ and $\psi \in \mathrm{Dom}(\hat T)$.
And, as the name suggests, a \emph{self-adjoint} operator $\hat T$ is one for which $\mathrm{Dom}(\hat T^*) = \mathrm{Dom}(\hat T)$ and $\hat T^* \psi = \hat T \psi$ for all $\psi \in \mathrm{Dom}(\hat T)$.

We also have the following proposition, which is important for our purposes. If
$\hat T$ is symmetric, then $\hat T^*$ is an extension of $\hat T$, which 
means that $\mathrm{Dom}(\hat T^*) \supset \mathrm{Dom}(\hat T)$ and $\hat T^* \psi = \hat T \psi$ for all $\psi \in \mathrm{Dom}(\hat T)$. (See e.g.\ Proposition 9.4 of Hall \cite{Hall}.)
So since we have already shown that $\mathrm{Dom}(\hat H_\mathrm{I})$ is dense in $\mathbf{H}$ and that $\hat H_\mathrm{I}$ is symmetric, we have therefore only left to show that $\mathrm{Dom}(\hat H_\mathrm{I}^*) \subset \mathrm{Dom}(\hat H_\mathrm{I})$ in order to show that $\hat H_\mathrm{I}$ is self-adjoint, as this will imply that $\mathrm{Dom}(\hat H_\mathrm{I}^*) = \mathrm{Dom}(\hat H_\mathrm{I})$.

In order to do this, let us consider the set, call it $W_\Sigma$, of all $\ket{\psi} \in \mathrm{Dom}(\hat H_\mathrm{I})$ of the form
\begin{equation}
\begin{aligned}
	\ket{\psi} =
		\sum_{m=0}^{M} \ket{\chi^{(m)}}
	=
		\sum_{m=0}^{M} \sum_{n=m}^{\infty}\ket{\chi^{(m)}_n}
\end{aligned}
\end{equation}
for some $M\in \mathbb{N}$, where $\ket{\chi^{(m)}} = \ket{\chi^{(m)}_m} + \ket{\chi^{(m)}_{m+2}} + \ket{\chi^{(m)}_{m+4}} + \ldots \in W$ for all $m\in\{0,\ldots,M\}$.
Let us also for each $m$ let $D^{(m)}_{m+1}$ be the corresponding $D_{m+1}$ set for $\ket{\chi^{(m)}}$ that is part of its free parameters. 
We can then recall from Eq.\ (\ref{hat_A_chi_v1}) that we have
\begin{equation}
\begin{aligned}
	\ket{\hat A \chi^{(m)}} &=
		\ket{\hat A^-_m \chi^{(m)}_m} +
		\sum_{n=m+2}^{\infty} \ket{\hat A^-_{n,n} \chi^{(m)}_{n} + \hat A^+_{n-2} \chi^{(m)}_{n-2}} +
		\sum_{n=m+2}^{\infty} \sum_{l=1}^{n-1} \ket{\hat A^-_{n,l} \chi^{(m)}_{n}}
	\\&=
		\ket{\hat A^-_m \chi^{(m)}_m} +
		\ket{1_{D^{(m)}_{m+1} \times \mathbb{R}^3}\hat A^+_m \chi^{(m)}_m} +
		\sum_{n=m+2}^{\infty} \sum_{l=1}^{n-1} \ket{\hat A^-_{n,l} \chi^{(m)}_{n}},
	\label{hat_A_chi_v2}
\end{aligned}
\end{equation}
for each of these $\ket{\chi^{(m)}}$,
where we have also used Eq.\ (\ref{that_term}) to get the second equality.
And from Eq.\ (\ref{chi_norm_bound_simult}), we also get that
\begin{equation}
\begin{aligned}
	\|\psi\|^2 \leq
		C_2 \sum_{m=0}^\infty \|\chi^{(m)}_m \|^2
	=
		C_2  \Big\|\sum_{m=0}^\infty \ket{\chi^{(m)}_m} \Big\|^2
\end{aligned}
\end{equation}
for some $C_2 > 0$.

Now, if we look at the combined contribution to $\braket{\phi | \hat A \psi}\!/\|\psi\|$, $\ket{\psi} \in W_\Sigma$, $\|\psi\| > 0$, coming from the last term on the right-hand side of Eq.\ (\ref{hat_A_chi_v2}) for any $\ket{\phi} \in \mathbf{H}$, we see that this is bounded by
\begin{equation}
\begin{aligned}
	\frac{1}{\|\psi\|}
	\sum_{m=0}^{M} \sum_{n=m+2}^{\infty} \sum_{l=1}^{n-1}
		\braket{\phi | \hat A^-_{n,l} \chi^{(m)}_{n}}
	&\leq
		\frac{
			\sum_{m=0}^{M}
			\|\phi\|
			\big\| \sum_{n=m+2}^{\infty} \sum_{l=1}^{n-1} \hat A^-_{n,l} \chi_{n}\big\|
		}{
			\sum_{m=0}^M \|\chi^{(m)}_m\|
		}
	\leq
		\sqrt{C_1} \|\phi\|
\end{aligned}
\end{equation}
for some $C_1 > 0$, where we have thus also used Eq.\ (\ref{A_l_less_than_n_chi_n_j_norm_bound_simult}) to get the last equality.
This shows us that $\psi \mapsto \braket{\phi | \hat A \psi}\!/\|\psi\|$, $\ket{\psi} \in \mathrm{Dom}(\hat H_\mathrm{I})$, $\|\psi\| > 0$, is bounded only if
\begin{equation}
\begin{aligned}
	\psi \mapsto
		\frac{\braket{\phi | \psi'}}{\|\psi\|},
	\quad
	\ket{\psi} \in W_\Sigma,
	\quad
	\|\psi\| > 0,
	\label{functional_for_W_Sigma}
\end{aligned}
\end{equation}
is bounded, where we define $\ket{\psi'} = \ket{\hat A \psi} - \sum_{n=m+2}^{\infty} \sum_{m=0}^{M} \sum_{l=1}^{n-1} \ket{\hat A^-_{n,l} \chi^{(m)}_{n}}$, such that
\begin{equation}
\begin{aligned}
	\braket{\phi | \psi'} =
		\sum_{m=1}^{M} 
			\braket{\phi_{m-1} | \hat A^-_m \chi^{(m)}_m}
		+
		\sum_{m=0}^{\infty}
			\braket{
				\phi_{m+1} |
				1_{D^{(m)}_{m+1} \times \mathbb{R}^3}\hat A^+_m \chi^{(m)}_{m}
			}.
	\label{psi_prime_braket}
\end{aligned}
\end{equation}

In order to then first of all show that $\mathrm{Dom}(\hat H_\mathrm{I}^*) \subset U$, let us pick an $M$ and a $\ket{\psi} \in W_\Sigma$ for which
\begin{equation}
\begin{aligned}
	\chi^{(m)}_m = \hat A^{R+}_{m-1} \phi_{m-1} + \hat A^{R-}_{m+1} \phi_{m+1}
	\label{chi_m_m_v1}
\end{aligned}
\end{equation}
for all $m\in\{0,\ldots, M\}$,
where we take $\hat A^{R+}_{-1} \phi_{-1}$ to be $0$. Let us here use similar definitions for $A^{R+}$ and $A^{R-}$ as in Eq.\ (\ref{cut_off_hat_As}), but let us use $1/R^a$, $a\geq1$, instead of $1/R$ for the lower cutoff, such that
\begin{equation}
\begin{aligned}
	\hat A^{R+}_{n-1} \psi_{n-1} &=
		1_{(B_{R,3} \setminus B_{1/R^a,3})^{n} \times \mathbb{R}^3} \hat A^{+}_{n-1} \psi_{n-1},
	\\
	\hat A^{R-}_{n} \psi_n &=
		1_{(B_{R,3} \setminus B_{1/R^a,3})^{n-1} \times \mathbb{R}^3} \hat A^{-}_{n}
			(1_{(B_{R,3} \setminus B_{1/R^a,3})^{n} \times \mathbb{R}^3} \psi_n).
	\label{cut_off_hat_As_v2}
\end{aligned}
\end{equation}
Since the support of each $\chi^{(m)}_m$ is therefore bounded by $R$ with respect to the $\mathbf{k}$-parameters, we can for each $m$ choose $D^{(m)}_{m+1}$ to be equal to $B_{R_m,3}^{m+1} \supset F_{m+1}$ for some sufficiently large $R_m > R$, similarly to what we did in Section \ref{sect_density}.
And since $A^{R-}$ and $A^{R+}$ are bounded, we then first of all get that $\|\chi^{(m)}_m\|, \|1_{D^{(m)}_{m+1} \times \mathbb{R}^3} \hat A^+_m \chi^{(m)}_{m}\| < \infty$.
We can furthermore see that
\begin{equation}
\begin{aligned}
	 \hat A^-_m \chi^{(m)}_m =
		\hat A^{R_m-}_m (\hat A^{R+}_{m-1} \phi_{m-1} + \hat A^{R-}_{m+1} \phi_{m+1})
	=
		\hat A^{R_m-}_m \chi^{(m)}_m,
\end{aligned}
\end{equation}
where we have thus used the fact that
$\hat A^-_m\hat A^{R+}_{m- 1} =
	\hat A^{R_m-}_m \hat A^{R+}_{m- 1}
$
and
$\hat A^-_m\hat A^{R-}_{m+ 1} =
	\hat A^{R_m-}_m \hat A^{R-}_{m+ 1}
$
when $R_m \geq R$.
Thus, we also get that $\|\hat A^-_m \chi^{(m)}_m\| < \infty$, which confirms that $\ket{\psi}$ exists as a member of $W_\Sigma$.

Next we note that since
$
	B_{R_m,3}^{m+1} \setminus (B_{R_m,3} \setminus B_{1/R_m^a,3})^{m+1} =
		B_{R_m,3}^{m+1} \setminus (B_{1/R_m^a,3}^\complement)^{m+1}
$,
we have
\begin{equation}
\begin{aligned}
	1_{D^{(m)}_{m+1} \times \mathbb{R}^3} \hat A^+_m \chi^{(m)}_{m} &=
		1_{B_{R_m,3}^{m+1} \times \mathbb{R}^3} \hat A^{+}_m
			\chi^{(m)}_{m}
	=
		\hat A^{R_m+}_m \chi^{(m)}_{m} +
		1_{(B_{R_m,3}^{m+1} \setminus (B_{1/R_m^a,3}^\complement)^{m+1}) \times \mathbb{R}^3}
		\hat A^{+}_m \chi^{(m)}_{m}.
\end{aligned}
\end{equation}
Since we can see that this last term tends to a zero vector when $a \to \infty$, let us therefore also define $\ket{\psi''} = \ket{\psi'} - \sum_{m=0}^{M} \sum_{l=1}^{n-1} \ket{1_{(B_{R_m,3}^{m+1} \setminus (B_{1/R_m^a,3}^\complement)^{m+1}) \times \mathbb{R}^3} \hat A^{+}_m \chi^{(m)}_{m}}$ to remove this term for now.
Then for 
$\braket{\phi | \psi''}$,
we get that
\begin{equation}
\begin{aligned}
	\braket{\phi | \psi''} &=
		\sum_{m=1}^{M} \braket{
			\phi_{m-1} |
			\hat A^{R_m-}_m \chi^{(m)}_m
		}
	+
		\sum_{m=0}^{M} \braket{
			\phi_{m+1} |
			\hat A^{R_m+}_m \chi^{(m)}_m
		}
	\\&=
		\sum_{m=1}^{M} \braket{
			\hat A^{R_m+}_{m-1} \phi_{m-1} |
			\chi^{(m)}_m
		}
	+
		\sum_{m=0}^{M} \braket{
			\hat A^{R_m-}_{m+1} \phi_{m+1} |
			\chi^{(m)}_m
		},
	\label{functional_calc_v1}
\end{aligned}
\end{equation}
where we for the last equality here have used the result of Eq.\ (\ref{cut_off_symmetry}).
And if we take $\hat A^{R_m+}_{-1} \phi_{-1}$ to be 0, and use Eq.\ (\ref{chi_m_m_v1}), we thus get that
\begin{equation}
\begin{aligned}
	\frac{\braket{\phi | \psi''}}{\|\psi\|} =
		\frac{1}{\|\psi\|}
		\sum_{m=0}^{M} \braket{
			\hat A^{R_m+}_{m-1} \phi_{m-1} +
			\hat A^{R_m-}_{m+1} \phi_{m+1}
			|
			\hat A^{R+}_{m-1} \phi_{m-1} +
			\hat A^{R-}_{m+1} \phi_{m+1}
		}.
	\label{functional_calc_v2}
\end{aligned}
\end{equation}

We can then note that in order for $\phi$ to be a member of $\mathrm{Dom}(\hat H_\mathrm{I}^*)$, the right-hand side of Eq.\ (\ref{functional_calc_v2}) has to be bounded when we let $a$ and each $R_m$ tend to infinity, as $\braket{\phi | \psi''}$ converges to $\braket{\phi | \psi'}$ in this limit. And it also has to be bounded if we then subsequently let $R, M\to \infty$ as well.
We must therefore have
\begin{equation}
\begin{aligned}
	\infty >
		\lim_{R, M \to\infty} \lim_{R_0, \ldots, R_M, a \to\infty}
			\frac{\braket{\phi | \psi''}}{\|\psi\|}
	\geq
		\lim_{R\to\infty}
		\frac{
			\sum_{m=0}^{\infty}
			\braket{(\hat A \phi)_m | (\hat A^R \phi)_m}
		}{
			\sqrt{C_2} \big\|\sum_{m=0}^\infty \ket{\chi^{(m)}_m} \big\|
		}
	=
		\frac{\|\hat A \phi\|^2}{\sqrt{C_2} \|\hat A \phi\|}
	=
		\frac{\|\hat A \phi\|}{\sqrt{C_2}}
	\label{functional_calc_v3}
\end{aligned}
\end{equation}
for all $\phi \in \mathrm{Dom}(\hat H_\mathrm{I}^*)$, where we have also used Eq.\ (\ref{chi_m_m_v1}) to recognize $\chi^{(m)}_m$ as $(\hat A^R \phi)_m$ to get the third equality.
This shows us that $\|\hat A \phi\| < \infty$ for all $\phi \in \mathrm{Dom}(\hat H_\mathrm{I}^*)$, and we therefore get that $\mathrm{Dom}(\hat H_\mathrm{I}^*) \subset U$.

To then show that $\mathrm{Dom}(\hat H_\mathrm{I}^*) \subset V$, we will use a similar strategy, but choose $D^{(m)}_{m+1} = F_{m+1}$ and
\begin{equation}
\begin{aligned}
	\chi^{(m)}_m =
		1_{F_{m} \times \mathbb{R}^3}
			\hat A^{R+}_{m-1} 1_{F_{m-1}^\complement \times \mathbb{R}^3} \phi_{m-1}
	\label{chi_m_m_v2}
\end{aligned}
\end{equation}
instead for all $m\in\{2,\ldots,M\}$. We will also choose $\chi^{(0)}_0$ and $\chi^{(1)}_1$ to be zero everywhere. We then first of all note that
$\mathrm{supp}(\chi^{(m)}_m) \subset F_m \times \mathbb{R}^3$,
which means that
\begin{equation}
\begin{aligned}
	\mathrm{supp}(\hat A^+_m \chi^{(m)}_m) \subset
		\big( \mathcal{S} (F_{m} \times \mathbb{R}^3) \big) \times \mathbb{R}^3.
\end{aligned}
\end{equation}
And since we have from Eq.\ (\ref{F_n_minus_1_productions_are_in_F_n_complement_v3}) that\begin{equation}
	\mathcal{S} (F_{m} \times \mathbb{R}^3) \subset
		F_{m+1}^\complement
	\label{F_n_minus_1_productions_are_in_F_n_complement_v4}
\end{equation}
for all $m\geq 2$,
we thus get that
\begin{equation}
\begin{aligned}
	\mathrm{supp}(\hat A^+_m \chi^{(m)}_m) \subset
		F_{m+1}^\complement \times \mathbb{R}^3.
\end{aligned}
\end{equation}
So for our $1_{D^{(m)}_{m+1} \times \mathbb{R}^3} \hat A^+_m \chi^{(m)}_{m}$ function with $D^{(m)}_{m+1} = F_{m+1}$, we therefore get that
\begin{equation}
\begin{aligned}
	\mathrm{supp}(1_{D^{(m)}_{m+1} \times \mathbb{R}^3} \hat A^+_m \chi^{(m)}_{m}) \subset
		(F_{m+1} \cap F_{m+1}^\complement) \times \mathbb{R}^3
	=
		\emptyset.
\end{aligned}
\end{equation}
This means that $\ket{1_{D^{(m)}_{m+1} \times \mathbb{R}^3} \hat A^+_m \chi^{(m)}_{m}}$ is just a zero vector for each $m$, and for $\braket{\phi | \psi'}$ of Eq.\ (\ref{psi_prime_braket}), we thus get that
\begin{equation}
\begin{aligned}
	\braket{\phi | \psi'} =
		\sum_{m=2}^{M} 
			\braket{\phi_{m-1} | \hat A^-_m \chi^{(m)}_m}
	=
		\sum_{m=2}^{M} \braket{
			\phi_{m-1} |
			\hat A^-_m 1_{F_{m} \times \mathbb{R}^3}
				\hat A^{R+}_{m-1} 1_{F_{m-1}^\complement \times \mathbb{R}^3} \phi_{m-1}
		}.
	\label{psi_prime_braket_v2}
\end{aligned}
\end{equation}
We can then use the fact that $\hat A^-_m 1_{F_{m} \times \mathbb{R}^3} \hat A^{R+}_{m-1} = \hat A^{R_{m-1}-}_m 1_{F_{m} \times \mathbb{R}^3} \hat A^{R+}_{m-1}$ for sufficiently large $R_0, \ldots, R_M$, to further obtain
\begin{equation}
\begin{aligned}
	\braket{\phi | \psi'} &=
		\sum_{m=2}^{M} \braket{
			\hat A^{R_{m-1}+}_{m-1} \phi_{m-1} |
			1_{F_{m} \times \mathbb{R}^3}
				\hat A^{R+}_{m-1} 1_{F_{m-1}^\complement \times \mathbb{R}^3} \phi_{m-1}
		}
	\\&=
		\sum_{m=2}^{M} \braket{
			1_{F_{m} \times \mathbb{R}^3} \hat A^{R_{m-1}+}_{m-1} \phi_{m-1} |
			1_{F_{m} \times \mathbb{R}^3}
				\hat A^{R+}_{m-1} 1_{F_{m-1}^\complement \times \mathbb{R}^3} \phi_{m-1}
		}.
	\label{psi_prime_braket_v3}
\end{aligned}
\end{equation}
Let us then split each $\phi_{m-1}$ up into
$
	\phi_{m-1} =
		1_{F_{m-1}^\complement \times \mathbb{R}^3} \phi_{m-1} +
		1_{F_{m-1} \times \mathbb{R}^3} \phi_{m-1}
$
for each $m$, and note that
\begin{equation}
\begin{aligned}
	\mathrm{supp}(\hat A^{R_{m-1}+}_{m-1} 1_{F_{m-1} \times \mathbb{R}^3} \phi_{m-1}) \subset
		\big( \mathcal{S} (F_{m-1} \times \mathbb{R}^3) \big) \times \mathbb{R}^3
	\subset
		F_{m}^\complement \times \mathbb{R}^3,
\end{aligned}
\end{equation}
where we have used Eq.\ (\ref{F_n_minus_1_productions_are_in_F_n_complement_v4}) once again. Therefore $1_{F_{m} \times \mathbb{R}^3} \hat A^{R_{m-1}+}_{m-1} 1_{F_{m-1} \times \mathbb{R}^3} \phi_{m-1}$ must be zero everywhere, which then gives us that
\begin{equation}
\begin{aligned}
	\braket{\phi | \psi'} =
		\sum_{m=2}^{M} \braket{
			1_{F_{m} \times \mathbb{R}^3} \hat A^{R_{m-1}+}_{m-1}
				1_{F_{m-1}^\complement \times \mathbb{R}^3} \phi_{m-1} |
			1_{F_{m} \times \mathbb{R}^3} \hat A^{R+}_{m-1}
				1_{F_{m-1}^\complement \times \mathbb{R}^3} \phi_{m-1}
		}.
	\label{psi_prime_braket_v4}
\end{aligned}
\end{equation}

We can then repeat the process of letting each $R_m\to\infty$, followed by $R, M\to\infty$. (We can keep $a$ constant in this case.)
This then gives us that
\begin{equation}
\begin{aligned}
	\lim_{R, M\to\infty} \lim_{R_0, \ldots, R_M \to\infty}
		\frac{\braket{\phi | \psi'}}{\|\psi\|}
	\geq
		\frac{
			\sum_{m=2}^{\infty}
			\|
				1_{F_{m} \times \mathbb{R}^3} \hat A^{+}_{m-1}
					1_{F_{m-1}^\complement \times \mathbb{R}^3} \phi_{m-1}
			\|^2
		}{
			\sqrt{C_2}\, 
			\big\|\!
				\sum_{m=2}^{\infty} \ket{
					1_{F_{m} \times \mathbb{R}^3} \hat A^{+}_{m-1}
						1_{F_{m-1}^\complement \times \mathbb{R}^3} \phi_{m-1}
				}\!
			\big\|
		}
	=
		\frac{\| \hat 1_{T} \hat A^+ \hat 1_{T^\complement} \phi \|}{\sqrt{C_2}}
	\label{functional_calc_v4}
\end{aligned}
\end{equation}
must be finite for all $\phi \in \mathrm{Dom}(\hat H_\mathrm{I}^*)$, where we have used Eq.\ (\ref{hat_1_Ts}) 
for the last equality.
We thus obtain that $\| \hat 1_{T} \hat A^+ \hat 1_{T^\complement} \phi \| < \infty$ for all $\phi \in \mathrm{Dom}(\hat H_\mathrm{I}^*)$, and we can therefore also conclude that $\mathrm{Dom}(\hat H_\mathrm{I}^*) \subset V$.

This completes the proof as we have now shown that $\mathrm{Dom}(\hat H_\mathrm{I}^*) \subset U \cap V = \mathrm{Dom}(\hat H_\mathrm{I})$. And since we already have that $\mathrm{Dom}(\hat H_\mathrm{I}) \subset \mathrm{Dom}(\hat H_\mathrm{I}^*)$ due to $\hat H_\mathrm{I}$ being symmetric, we thus get that $\mathrm{Dom}(\hat H_\mathrm{I}^*) = \mathrm{Dom}(\hat H_\mathrm{I})$. We also already have that $\hat H_\mathrm{I}^* \psi = \hat H_\mathrm{I} \psi$ for all $\psi \in \mathrm{Dom}(\hat H_\mathrm{I})$, and we can therefore conclude that $\hat H_\mathrm{I}$ is self-adjoint.
And since we have thus shown both that $\mathrm{Dom}(\hat H_\mathrm{I})$ is dense in $\mathbf{H}$ and that $\hat H_\mathrm{I}$ is self-adjoint, we hereby obtain Proposition \ref{prop_1}.

\section{Conclusion}

We have defined a simplified version of the Dirac interaction operator, $\hat H_\mathrm{I}$, with no cutoffs on a certain domain. We have then shown, in Sections \ref{sect_W_subset_V}--\ref{sect_density}, that this domain is dense in the Hilbert space. In Section \ref{sect_sym}, we have furthermore shown that $\hat H_\mathrm{I}$ is symmetric on the domain. And finally, in Section \ref{sect_SA}, we have shown that $\hat H_\mathrm{I}$ is also self-adjoint on the domain, thus obtaining Proposition \ref{prop_1}.

\begin{appendices}

\section[Symmetrizing the Hilbert space]{Symmetrizing the Hilbert space} \label{app_symmetrize}

In this appendix, we will briefly discuss how to extend the result of this paper to a restricted Hilbert space that only includes vectors, $\ket{\psi}$, where each $\psi_n$ is symmetric with respect to the $\mathbf{k}$-coordinates.
For this, one can define a symmetrizing operator, call it $\hat{\mathcal{S}}$, given by $\hat{\mathcal{S}} \psi = (\psi_0, \hat{\mathcal{S}}_1 \psi_1, \hat{\mathcal{S}}_2 \psi_2, \ldots)$, with each $\hat{\mathcal{S}}_n$ given by
\begin{equation}
\begin{aligned}
	\hat{\mathcal{S}}_n \psi_n(\mathbf{k}_1,\ldots, \mathbf{k}_n; \mathbf{p}) =
		\frac{1}{n!} \big(
			\psi_n(\mathbf{k}_1,\ldots, \mathbf{k}_n; \mathbf{p}) +
			\text{all other permutations of $\mathbf{k}_1,\ldots, \mathbf{k}_n$}
		\big).
\end{aligned}
\end{equation}
It is easy to then show that $\hat{\mathcal{S}}$ is bounded from above by 1, and that it commutes with $\hat A$. One can also see that $\ket{\psi} \in \mathrm{Dom}(\hat H_\mathrm{I})$ implies $\ket{\hat{\mathcal{S}}\psi} \in \mathrm{Dom}(\hat H_\mathrm{I})$.

One way to proceed from there is then to show that 
$\braket{\hat{\mathcal{S}} \phi | \hat{\mathcal{S}} \psi} = \braket{\hat{\mathcal{S}} \phi | \psi}$ for all $\phi,\psi \in \mathbf{H}'$,
which will give us that $\braket{\hat{\mathcal{S}} \phi | \hat A \psi} = \braket{\hat{\mathcal{S}} \phi | \hat{\mathcal{S}} \hat A \psi} = \braket{\hat{\mathcal{S}} \phi | \hat A \hat{\mathcal{S}} \psi}$. This tells us that $\hat{\mathcal{S}} \phi$ is a member of the symmetrized version of $\mathrm{Dom}(\hat H_\mathrm{I}^*)$ if and only if it is also a member of the original $\mathrm{Dom}(\hat H_\mathrm{I}^*)$. And it follows that the symmetrized version of $\hat H_\mathrm{I}$, (on the symmetrized version of $\mathbf{H}$) is also self-adjoint.

\section{Adding a free energy to the operator}

Let us here briefly discuss how to potentially make the technique of this paper work when we add a free energy, $\hat H_0$, to the operator. This $\hat H_0$ should go asymptotically as $|\mathbf{k}_1| + \ldots + |\mathbf{k}_n| + |\mathbf{p}|$.
Being able to add this free energy is of course very important if we hope to extend this result to physically meaningful Hamiltonians.

Now, $\mathrm{Dom}(\hat H_\mathrm{I})$ and $W$ has actually been deliberately constructed in this paper such that $\hat H_0 \ket{\chi}$ also has a finite norm for all $\ket{\chi} \in W$ where the norm of $\hat H_0 \ket{\chi_m}$ is finite. We should thus still be able to show that $W$ is a subset of $U$ when we alter $U$ to be the subset of all $\ket{\psi} \in \mathbf{H}$ for which $(\hat H_0 + \hat H_\mathrm{I})\ket{\psi} \in \mathbf{H}$.

For the part of the proof in Section \ref{sect_SA}, which aims to show that $\hat H = \hat H_0 + \hat H_\mathrm{I}$ is self-adjoint, we can then see that we also ought to redefine $V$ in order to make this proof work, namely as the set of all $\ket{\psi}$ for which
\begin{equation}
\begin{aligned}
	\|
		\hat 1_{T} \hat A^+ \hat 1_{T^\complement} \psi +
		\hat 1_{T} \hat A^0 \psi
	\| < \infty,
	\label{V_def_v2}
\end{aligned}
\end{equation}
where we define $\ket{\hat A^0 \psi} = \hat H_0 \ket{\psi}$.

This, however, also means that the proof in Section \ref{sect_sym}, which aims to show that $\hat H$ is symmetric, has to change, and in particular the part that shows the lemma of Eq.\ (\ref{hat_J_norm_lower_bound_v2}). This particular task is not so trivial, but is should not be too hard, still. Intuitively, one can see that in order for $\hat H_0\ket{\psi}$ to interfere with the lower bound on $\| 1_{F_{N+1,n} \times \mathbb{R}^3} \hat A^+_{N,n} \psi'_{N, \mathbf i} \|$, so to speak, $\hat H_0 \ket{\psi_{N+1}}$ then has to cancel some of $\ket{1_{F_{N+1,n} \times \mathbb{R}^3} \hat A^+_{N,n} \psi'_{N, \mathbf i}}$. But in order for $\hat H_0 \ket{\psi_{N+1}}$ to cancel any significant portion of $\ket{1_{F_{N+1,n} \times \mathbb{R}^3} \hat A^+_{N,n} \psi'_{N, \mathbf i}}$, we can see that this will require $\psi_{N+1}(\mathbf{k}_1, \ldots, \mathbf{k}_{N+1})$ to go approximately as $1_{F_{N+1,n} \times \mathbb{R}^3}\psi_{N}(\mathbf{k}_1, \ldots, \widehat{\mathbf{k}_n}, \ldots, \mathbf{k}_{N+1})/|\mathbf{k}_n|^{3/2}$. And when we compute the norm of this, we see that it will go as $p'_{N+1}(i_1, \ldots, i_N)\|\psi'_{N, \mathbf i}\|/2$. Since $p'_{N+1}(i_1, \ldots, i_N)$ is a rapidly growing function, it will thus mean that $\|\psi_{N+1}\|$ has to continuously be much larger than $\|\psi'_{N, \mathbf i}\|$ in order to interfere with our bound, which should not be possible if $\|\psi\|$ 
converges.

Thus, we should be able to change the proof of Eq.\ (\ref{hat_J_norm_lower_bound_v2}) to accommodate for the altered $V$ when we add $\hat H_0$ to the operator. And since this part seems to be the only one that is potentially somewhat tricky, at least at an overall glance, we can therefore see that we might very well be able to extend Proposition \ref{prop_1} to such
operators that include a free energy.

\section{Handling the perturbed vacuum}

A big hope is
that the technique
of this paper
can also at some point be extended to a theory such as the full theory of quantum electrodynamics (QED).
One of the key tasks for doing this, as can be seen e.g.\ in Damgaard \cite{Damgaard_QED}, is to find a way to handle the infinite vacuum fluctuations that seem to appear when negative-energy fermion solutions are reinterpreted as antiparticles. However, there might be a solution to handle this perturbed vacuum if one can only show that the vacuum-perturbing part of the Hamiltonian becomes self-adjoint if it is multiplied by factors of $1/\sqrt{\mathcal V}$, with $\mathcal V$ being volume of the system, before going to the continuum limit. If this can be achieved, the idea is then to use this to show that the $\varepsilon$-almost eigenstates of the perturbed vacuum converge in the continuum limit. And from there it seems that one might be able to show that the transition amplitudes of the interaction between such a ``vacuum state'' and any ``physical state'' will vanish in this limit.\footnote{
	And even if the transition amplitudes do not vanish, perturbation theory still tells us that the growing energy of such vacuum solutions in the continuum limit,
	which comes when we put back the factors of $\sqrt{\mathcal V}$,
	ought to make the interaction vanish anyway between any physical state and any vacuum 
	state
	that does not have a vanishing energy to begin with.
}
This would then tell us that the perturbed vacuum does not interact with the physical particles of the system, and that we can simply remove the vacuum-perturbing terms in the Hamiltonian.
Needless to say, this would be an incredible prospect.
However, it requires us to be able to prove the self-adjointness of such operators first.

\end{appendices}



\begin{thebibliography}{20}


\bibitem{Hall}
	B.\ C.\ Hall, 
	\textit{Quantum Theory for Mathematicians} 
	(Springer, New York, 2013).



\bibitem{Reed_and_Simon}
	M.\ Reed and B.\ Simon,
	\textit{Methods of Modern Mathematical Physics}, 
	Vol.\ II: Fourier Analysis, Self-adjointness
	(Academic Press, New York, 1975).


\bibitem{Weinberg}
	S.\ Weinberg, 
	\textit{The Quantum Theory of Fields},
	Vol.\ I: Foundations
	(Cambridge University Press, Cambridge, 1995).







\bibitem{Srednicki}
	M.\ Srednicki,
	\textit{Quantum Field Theory}
	(Cambridge University Press, Cambridge, 2007).


\bibitem{L_and_B}
	T.\ Lancaster and S.\ J.\ Blundell,
	\textit{Quantum Field Theory for the Gifted Amateur}
	(Oxford University Press, Oxford, 2014).



\bibitem{Damgaard_QED}
	M.\ J.\ Damgaard,
	\textit{An alternative derivation of the Hamiltonian of quantum electrodynamics},
	viXra:2210.0044v2 [quant-ph].




%
%
%
%
%
%
%
%
%
%
%
%
%
%
%
%
%
%







\end{thebibliography}
\end{document}